\documentclass[twocolumn]{aastex631}

\usepackage{amssymb}
\usepackage{mathrsfs}
\usepackage{amsmath}

\begin{document}

\title{Recovery of High-energy Low-frequency Quasi-periodic Oscillations from Black Hole X-ray Binary MAXI J1535--571 with a Hilbert-Huang Transform Method}

\correspondingauthor{Qing C. Shui}
\email{shuiqc@ihep.ac.cn}
\correspondingauthor{S. Zhang}
\email{szhang@ihep.ac.cn}
\correspondingauthor{Shuang N. Zhang}
\email{zhangsn@ihep.ac.cn}

\author[0000-0001-5160-3344]{Qing C. Shui}
\affiliation{Key Laboratory of Particle Astrophysics, Institute of High Energy Physics, Chinese Academy of Sciences, 100049, Beijing, China}
\affiliation{University of Chinese Academy of Sciences, Chinese Academy of Sciences, 100049, Beijing, China}

\author{S. Zhang}
\affiliation{Key Laboratory of Particle Astrophysics, Institute of High Energy Physics, Chinese Academy of Sciences, 100049, Beijing, China}

\author[0000-0001-5586-1017]{Shuang N. Zhang}
\affiliation{Key Laboratory of Particle Astrophysics, Institute of High Energy Physics, Chinese Academy of Sciences, 100049, Beijing, China}
\affiliation{University of Chinese Academy of Sciences, Chinese Academy of Sciences, 100049, Beijing, China}

\author{Yu P. Chen}
\affiliation{Key Laboratory of Particle Astrophysics, Institute of High Energy Physics, Chinese Academy of Sciences, 100049, Beijing, China}

\author[0000-0003-3188-9079]{Ling D. Kong}
\affiliation{Institut f\"{u}r Astronomie und Astrophysik, Kepler Center for Astro and Particle Physics, Eberhard Karls, Universit\"{a}t, Sand 1, D-72076 T\"{u}bingen, Germany}

\author[0000-0002-5554-1088]{Jing Q. Peng}
\affiliation{Key Laboratory of Particle Astrophysics, Institute of High Energy Physics, Chinese Academy of Sciences, 100049, Beijing, China}
\affiliation{University of Chinese Academy of Sciences, Chinese Academy of Sciences, 100049, Beijing, China}

\author{L. Ji}
\affiliation{School of Physics and Astronomy, Sun Yat-Sen University, Zhuhai, 519082, China}

\author[0000-0002-6454-9540]{Peng J. Wang}
\affiliation{Key Laboratory of Particle Astrophysics, Institute of High Energy Physics, Chinese Academy of Sciences, 100049, Beijing, China}
\affiliation{University of Chinese Academy of Sciences, Chinese Academy of Sciences, 100049, Beijing, China}

\author[0000-0003-4856-2275]{Z. Chang}
\affiliation{Key Laboratory of Particle Astrophysics, Institute of High Energy Physics, Chinese Academy of Sciences, 100049, Beijing, China}

\author{Zhuo L. Yu}
\affiliation{Key Laboratory of Particle Astrophysics, Institute of High Energy Physics, Chinese Academy of Sciences, 100049, Beijing, China}

\author[0000-0002-0638-088X]{Hong X. Yin}
\affiliation{Shandong Key Laboratory of Optical Astronomy and Solar-Terrestrial Environment, School of Space Science and Physics, Institute of Space Sciences, Shandong University, Weihai, Shandong 264209, China}

\author[0000-0002-9796-2585]{Jin L. Qu}
\affiliation{Key Laboratory of Particle Astrophysics, Institute of High Energy Physics, Chinese Academy of Sciences, 100049, Beijing, China}

\author[0000-0002-2705-4338]{L. Tao}
\affiliation{Key Laboratory of Particle Astrophysics, Institute of High Energy Physics, Chinese Academy of Sciences, 100049, Beijing, China}

\author[0000-0002-2749-6638]{Ming Y. Ge}
\affiliation{Key Laboratory of Particle Astrophysics, Institute of High Energy Physics, Chinese Academy of Sciences, 100049, Beijing, China}

\author[0000-0002-2032-2440]{X. Ma}
\affiliation{Key Laboratory of Particle Astrophysics, Institute of High Energy Physics, Chinese Academy of Sciences, 100049, Beijing, China}

\author{L. Zhang}
\affiliation{Key Laboratory of Particle Astrophysics, Institute of High Energy Physics, Chinese Academy of Sciences, 100049, Beijing, China}



\author{W. Yu}
\affiliation{Key Laboratory of Particle Astrophysics, Institute of High Energy Physics, Chinese Academy of Sciences, 100049, Beijing, China}


\author{J. Li}
\affiliation{CAS Key Laboratory for Research in Galaxies and Cosmology, Department of Astronomy, University of Science and Technology of China, Hefei 230026, China}
\begin{abstract}
We propose a method based on the Hilbert-Huang transform (HHT) to recover the high-energy waveform of low-frequency quasi-periodic oscillations (LFQPOs). Based on the method, we successfully obtain the modulation of the phase-folded light curve above 170 keV using the QPO phase reconstructed at lower energies in MAXI J1535--571 with Insight-HXMT observations. A comprehensive simulation study is conducted to demonstrate that such modulation indeed originates from the QPO. Thus the highest energies turn out to significantly exceed the upper limit of $\sim$100 keV for QPOs reported previously using the Fourier method, marking the first opportunity to study QPO properties above 100 keV in this source. Detailed analyses of these high-energy QPO profiles reveal different QPO properties between the 30--100 keV and 100--200 keV energy ranges: the phase lag remains relatively stable, and the amplitude slightly increases below $\sim100$ keV, whereas above this threshold, soft phase lags and a decrease in amplitude are observed. Given the reports of a hard tail detection in broad spectroscopy, we propose that the newly discovered QPO properties above 100 keV are dominated by the hard tail component, possibly stemming from a relativistic jet. Our findings also indicate a strong correlation between the QPOs originating from the jet and corona, supporting the scenario of jet-corona coupling precssion. We emphasize that our proposed HHT-based method can serve as an efficient manner in expanding the high energy band for studying QPOs, thereby enhancing our understanding of their origin.
\end{abstract}



\section{Introduction} \label{sec:intro}
Following a prolonged period of quiescence at low-flux levels, black hole X-ray binaries (BHXRBs) can experience outbursts, with substantial changes in their X-ray emission properties \citep{2006ARA&A..44...49R,2007A&ARv..15....1D}. During a complete outburst, the source typically exhibits four distinct canonical states: the low/hard state (LHS), hard intermediate state (HIMS), soft intermediate state (SIMS) and high soft state (HSS). A source exhibits distinct X-ray spectral and variability properties in different states \citep[][]{2005A&A...440..207B,2005Ap&SS.300..107H}.  Low frequency quasi-periodic oscillations (LFQPOs) are observed in most transient BHXRBs, typically ranging from 0.1 to 30 Hz \citep{1989ARA&A..27..517V}. These oscillations are characterized by a narrow peak with finite width in the power density spectra (PDS) and are typically classified into three types (A, B, and C) based on their centroid frequency, $Q$ factor (ratio of frequency to the full width at half maximum), and root-mean-square (rms) amplitude \citep{1999ApJ...526L..33W,2005ApJ...629..403C,2006ARA&A..44...49R}.

LFQPOs are thought to originate from the innermost part of the accretion flow, where the most X-ray emissions and fast variability are produced. Fast X-ray variability is observed across a wide range of timescales \citep{2014SSRv..183...43B}. In addition to the QPO component, broadband continuum components also contribute to the PDS, which represent fast aperiodic variability. Several models have been suggested to account for this noise component, such as the shot noise model, coronal flare model, and fluctuation propagation model \citep{1972ApJ...174L..35T,1981ApJ...246..494N,1993ApJ...411L..91S,1994ApJ...435L.125M,2011MNRAS.415.2323I}. For LFQPOs, various models have been also proposed, based on either the geometric effect or the intrinsic variability of the accretion flow. Intrinsic models include the trapped corrugation modes \citep{1990PASJ...42...99K,1999PhR...311..259W}, the Accretion-ejection instability model \citep[AEI,][]{1999A&A...349.1003T}, and the Two-Component Advection Flow model \citep{1996ApJ...457..805M}, etc. Most geometric models, on the other hand, are associated with the relativistic Lense-Thirring (L-T) precession effect, such as the relativistic precession model \citep[RPM;][]{1999ApJ...524L..63S}, L-T precession of the entire hot accretion flow \citep{2009MNRAS.397L.101I} and precession of the jet \citep{2016MNRAS.460.2796S,2019MNRAS.485.3834D,2021NatAs...5...94M,2023ApJ...948..116M}.

Componization is widely considered to be the radiative mechanism behind type-C QPOs \citep[e.g.,][]{2022NatAs...6..577M,2022MNRAS.515.2099B}, because observational studies show that the factional rms of type-C QPOs generally increases with photon energy \citep{2017ApJ...845..143Z,2018ApJ...866..122H,2020JHEAp..25...29K,2020MNRAS.494.1375Z} and no evident disc-like component is detected in their rms spectra \citep{2006MNRAS.370..405S,2014MNRAS.438..657A,2016MNRAS.458.1778A,2023ApJ...957...84S}. Furthermore, \citet{2021NatAs...5...94M} reported the discovery of tpye-C QPOs above 200 keV, further indicating a strong association with Comptonized emission. Additionally, inclination-dependent amplitudes and time lags of type-C QPOs \citep[see][]{2015MNRAS.447.2059M,2017MNRAS.464.2643V}, and QPO phase-dependent reflection component revealed with phase-resolved spectroscopy \citep[see][]{2015MNRAS.446.3516I,2016MNRAS.461.1967I,2017MNRAS.464.2979I,2022MNRAS.511..255N,2023ApJ...957...84S}, may suggest type-C QPOs originate from some geometrical effects. For recent reviews of observations and theories of LFQPOs, the reader is referred to \citet{2019NewAR..8501524I}.

The study of LFQPOs in high energy bands ($>30$ keV) is essential to our understanding of their origin, since the radiative mechanism and geometric properties of LFQPOs could be further constrained \citep{2018ApJ...866..122H,2020JHEAp..25...29K,2021NatAs...5...94M,2021RAA....21...70L}. Thanks to the wide energy range of Insigt-HXMT, we can study high-energy LFQPOs above 100 keV in some BHXRBs in details \citep[see e.g.][]{2021NatAs...5...94M}. Furthermore, $\gamma$-ray QPOs have been detected by Fermi-LAT in numerous bright blazars \citep[see e.g.][]{2014ApJ...793L...1S,2015ApJ...813L..41A,2023MNRAS.523L..52B}. Consequently, certain BHXRBs may have the potential to exhibit $\gamma$-ray QPOs, which could involve different radiative mechanisms from those observed in the X-ray bands, and warrant further detailed investigation in the future. In this Letter, we propose a novel technique based on the Hilbert-Huang transform (HHT) for recovering high-energy QPO signals in MAXI J1535--571, which is one of the brightest BHXRBs observed by Insight-HXMT. The HHT is a powerful tool for analyzing phenomena with non-stationary periodicity proposed by \citet{1998RSPSA.454..903H} and has been successfully applied in astronomical researches \citep{2007PhRvD..75f1101C,2011ApJ...740...67H,2014ApJ...788...31H,2015ApJ...815...74S,2020ApJ...900..116H,2023ApJ...951..130Y,2023ApJ...957...84S}.The HHT allows us to track changes in the frequency of QPOs on short time-scales and conduct their phase-resolved spectral analysis.

MAXI J1535--571 is an X-ray transient discovered by Gas Slit Camera of the Monitor of All-sky X-ray Image (MAXI/GSC) and and the Burst Alert Telescope of Swift (Swift/BAT) independently when it went into outburst on 2017 September 2 \citep{2017ATel10700....1K,2017ATel10699....1N}. Subsequently, this outburst was extensively monitored at radio \citep{2019ApJ...878L..28P,2019ApJ...883..198R}, sub-millimetre \citep{2017ATel10745....1T}, infrared and optical \citep{2018ApJ...867..114B}, and X-ray \citep{2018ApJ...866..122H,2018PASJ...70...95N,2018ApJ...865L..15S,2018MNRAS.480.4443T,2019MNRAS.488..720B} wavelengths. From H\uppercase\expandafter{\romannumeral1} absorption, the source distance was estimated to be $4.1^{+0.6}_{-0.5}$kpc \citep{2019MNRAS.488L.129C}. LFQPOs (type-A, B and C) were detected by multiple X-ray instruments in the spectral state of LHS, HIMS and SIMS \citep{2018ApJ...866..122H,2018ApJ...868...71S,2018AstL...44..378M,2018ApJ...865L..15S}.

In this Letter, we propose a HHT-based technique to reconstruct the high-energy QPO waveforms. In Section~\ref{sec:2}, we outline the Insight-HXMT observations and data reductions. Then, in Section~\ref{sec:3}, we present a brief introduction to the HHT method and its application in recovering high-energy QPO signals, and then present the corresponding results derived from MAXI J1535--571. Finally, we discuss and summarize these results in Section~\ref{sec:4} and Section~\ref{sec:5}, respectively.

\section{Observations and Data Reduction} \label{sec:2}
Launched on June 15, 2017, the Hard X-ray Modulation Telescope (Insight-HXMT) is the first Chinese X-ray astronomy satellite \citep{2014SPIE.9144E..21Z,2020SCPMA..6349502Z}. The science payload of Insight-HXMT allows for observations across a broad energy band (1--250 keV) using three telescopes: the High Energy X-ray telescope (HE, composed of NaI/CsI, 20--250 keV), the Medium Energy X-ray telescope (ME, with a Si pin detector, 5--30 keV), and the Low Energy X-ray telescope (LE, using an SCD detector, 0.7--13 keV). For additional information about the Insight-HXMT mission, please refer to \citet{2019ApJ...879...61Z,2020SCPMA..6349503L,2020SCPMA..6349504C,2020SCPMA..6349505C}. 

\begin{figure}
\centering
    \includegraphics[width=\linewidth]{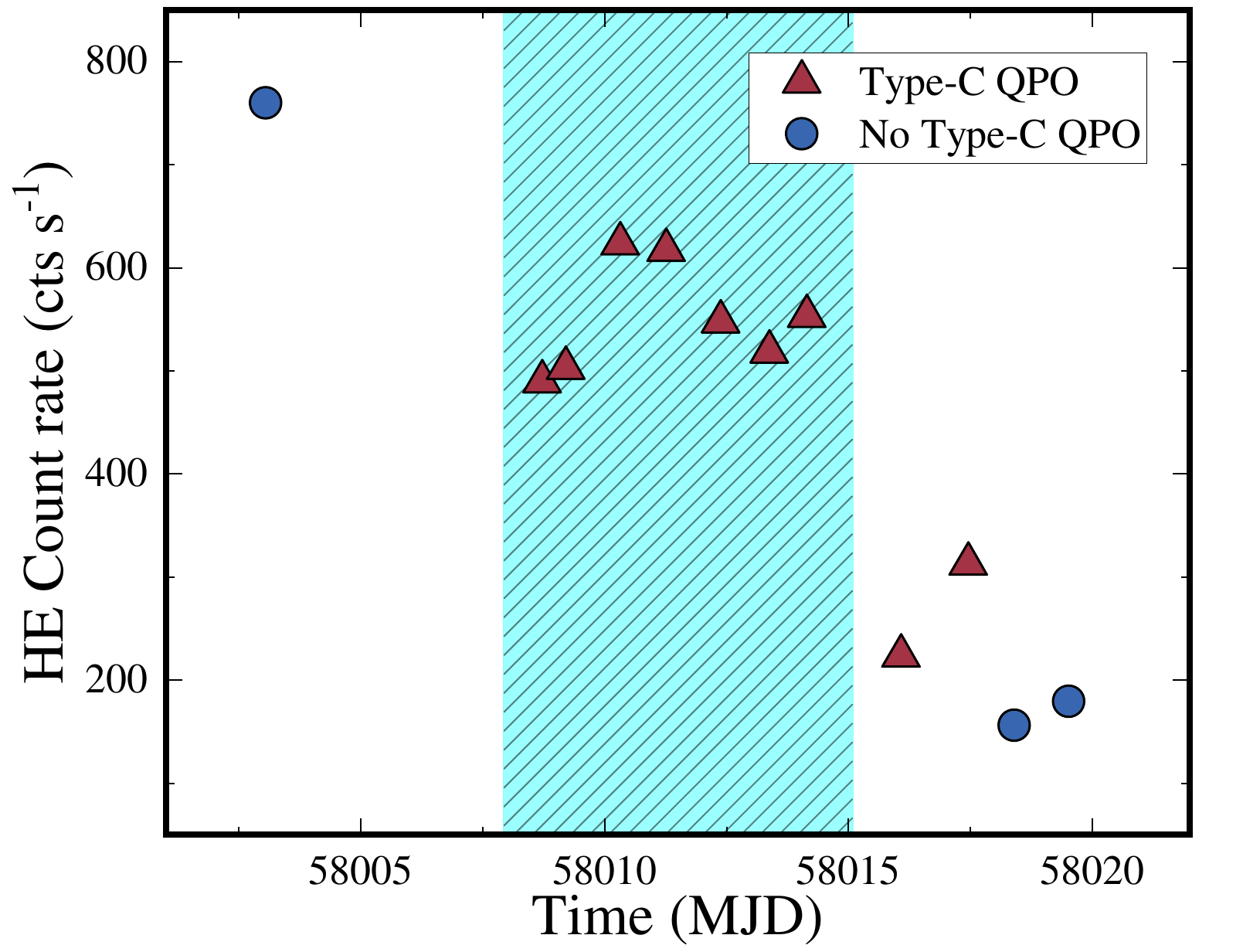}
    \caption{HE light curves of MAXI J1535--571. In this plot, red triangles denote observations with type-C QPO detection, while blue circles represent those without type-C QPO detection. Each point represents one observation, and the shaded area indicates the selected observations for detailed timing analyses.}
    \label{fig:1}
\end{figure}

We focus on high-energy signals of the type-C QPO detected by Insight-HXMT, so one of the bright systems MAXI J1535--571 with numerous type-C QPO observations \citep[see][]{2018ApJ...866..122H,2020JHEAp..25...29K,2022MNRAS.512.2686Z,2023ApJ...953..191Y} is investigated in this study. We extract the event data using Insight-HXMT Data Analysis Software \texttt{v2.05}\footnote{\url{http://hxmtweb.ihep.ac.cn/SoftDoc.jhtml}} and the current calibration model \texttt{v2.06}\footnote{\url{http://hxmtweb.ihep.ac.cn/software.jhtml}}, following the criteria recommended by the Insight-HXMT team: (1) elevation angle (ELV) greater than $10^{\circ}$; (2) geometric cutoff rigidity (COR) greater than 8 GeV; (3) pointing position offset less than $0.04^{\circ}$; and (4) good time intervals (GTIs) at least 300 s away from the South Atlantic Anomaly (SAA). Backgrounds are derived from blind detectors using the \texttt{LEBKGMAP}, \texttt{MEBKGMAP}, and \texttt{HEBKGMAP} tools, version 2.0.9, based on the standard Insight-HXMT background models \citep{2020JHEAp..27...14L,2020JHEAp..27...24L,2020JHEAp..27...44G}. All photons are corrected to the barycenter of the solar system for their arrival times using the \texttt{HXMTDAS} tool \texttt{hxbary}.
In Figure~\ref{fig:1}, we present the long-term net light curve for MAXI J1535--571 obtained from HE in the energy range of 30--120 keV. To ensure robust count statistics for HE, we select QPO observations with HE count rates in the energy range of 30--120 keV that closely approximate the highest level across the outburst (as indicated by the shaded area in Figure~\ref{fig:1}). We subsequently generate net light curves with a time resolution of 0.05 s for these selected observations, which are then subjected to the detailed timing analysis. Detailed information regarding the aforementioned observations is available in Table~\ref{tab:1}.

\begin{table}[]
    \centering
    \caption{Log of Insight-HXMT Observations Used in This Work\label{tab:1}}
    \begin{tabular}{ccccc}
    \hline
    \hline
    \# & ObsID & Start Time& QPO Frequency\\ 
    &   & (MJD) & (Hz)\\
    \hline
    1 & P011453500144 & 58008.48 & $2.53\pm0.02$ \\
    2 & P011453500145 & 58009.29 & $2.61\pm0.01$ \\
    3 & P011453500201 & 58010.32 & $1.78\pm0.01$ \\
    4 & P011453500301 & 58011.25 & $2.02\pm0.01$ \\
    5 & P011453500401 & 58012.38 & $2.76\pm0.01$ \\
    6 & P011453500501 & 58013.37 & $3.33\pm0.01$ \\
    7 & P011453500601 & 58014.14 & $3.32\pm0.03$ \\
    \hline
    \hline
    \end{tabular}
\end{table}

\section{Analysis and Results}
\label{sec:3}
The Fourier transform is a general tool for searching QPO signals in X-ray binaries. QPOs can be identified as a series of harmonic peaks with narrow widths in the PDS. Due to limited count statistics in the high-energy bands ($>100$ keV), previous studies typically combined multiple observations to achieve a higher signal-to-noise ratio (SNR) \citep[see e.g.,][]{2021NatAs...5...94M,2022ApJ...932....7Y}. Conversely, periodic folding is a common approach used in studies of neutron star pulsations to search for high-energy pulse profiles \citep[see e.g.,][]{2022ApJ...938..149H}. Since it has been demonstrated in BHXRBs that there exists an average underlying waveform of the LFQPO \citep{2015MNRAS.446.3516I,2019MNRAS.485.3834D}, similar folding methods could potentially reconstruct the high-energy QPO waveform if the high-energy QPO signal indeed exists. However, when dealing with QPO signals in black hole binaries, a simple folding of the light curve on a period is not suitable due to the non-deterministic evolution of their phases over time \citep[see e.g.,][]{2015MNRAS.446.3516I}. Specifically, \cite{1997ApJ...482..993M} observed the QPO phase to drift on a random walk away from that of a strictly periodic signal. There are several well-known methods for studying the time variations of the QPO frequency and amplitude, such as dynamic power spectra \citep[DPS, see e.g.][]{2020ApJ...891L..29H} and wavelet analysis \citep[see e.g.][]{2010A&A...515A..65L,2023A&A...677A.178Z}. In general, ``instantaneous" frequencies obtained from the DPS or wavelet analysis can be used for segmentally periodic folding the high-energy light curve to construct high-energy QPO waveforms. However, the DPS requires setting parameters for the width of the time window, and the wavelet analysis needs parameters for the decay length of the wavelet function, both of which significantly affect the performance of the methods. In addition to periodic folding, phase-folding is another approach for constructing the waveform for a non-stationary periodicity \citep[see e.g.][]{2015ApJ...815...74S,2023ApJ...957...84S}. Time-resolved complex power spectra can provide time variation of the phase information for the QPO, but it also faces challenges related to the selection of the width of the time window. Furthermore, it is still under debate how to separate the intrinsic QPO phase from the complex Fourier transform, as both QPO and noise components contribute to the Fourier transform at the QPO frequency \citep{2021NatAs...5...94M,2022MNRAS.515.1914Z}. The HHT is a powerful timing analysis tool that can provide an instantaneous phase function for the intrinsic QPO component at each time bin \citep[see e.g.][]{2015ApJ...815...74S,2023ApJ...957...84S}, effectively addressing the challenges of the aforementioned methods. In this section, we will first introduce a method for searching high-energy QPOs based on the HHT method.

\subsection{Hilbert-Huang Transform Analysis} 
\begin{figure*}
\centering
\includegraphics[width=0.9\textwidth]{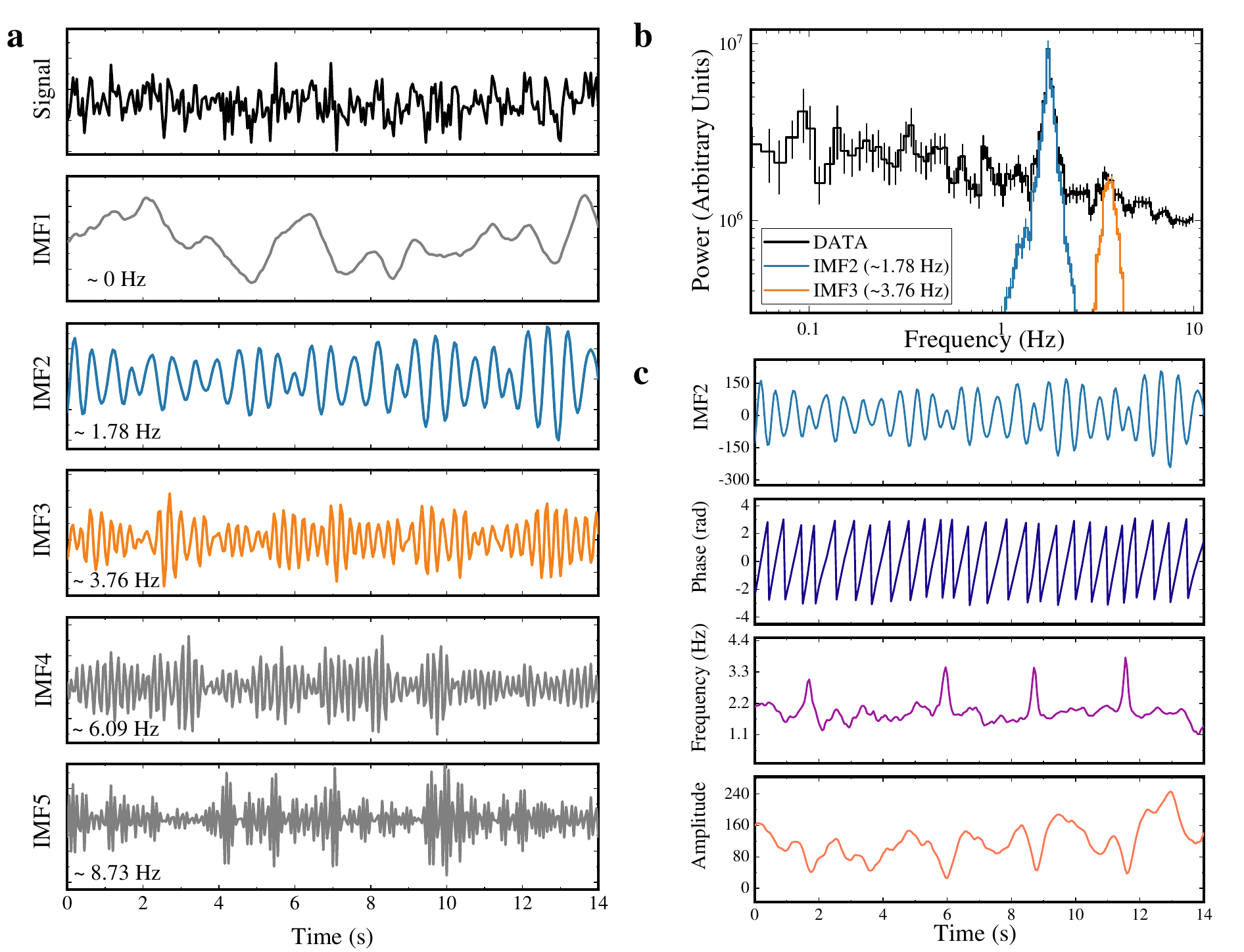}
    \caption{Hilbert-Huang transform analysis of a representative example of a 14-s-long light curve with a 0.05s time resolution of HE in energy range of 30--120 keV. \textbf{a}, representative example of a 14-s-long light curve and corresponding five IMFs obtained from VMD algorithm. \textbf{b},  power density spectra of the original HE light curve, as well as the second and third IMFs. \textbf{c}, the QPO (the second) IMF and corresponding phase function, instantaneous frequency and amplitude.}
    \label{fig:2}
\end{figure*}

The HHT, proposed by \citet{1998RSPSA.454..903H} as an adaptive data analysis method, offers a powerful tool for analyzing signals with non-stationary periodicity. It consists of two main components: mode decomposition, which aims to decomposing a signal into several intrinsic mode functions (IMFs), and Hilbert spectral analysis (HSA), which allows for the extraction of frequency and phase functions of the desired IMFs, such as the QPO component \citep[e.g.][]{2014ApJ...788...31H,2020ApJ...900..116H}. Following \citet{2023ApJ...957...84S}, we use the variational mode decomposition \citep[VMD,][]{6655981} method to extract the IMF correspond to the QPO component. Compared to the traditional method empirical mode decomposition \citep[EMD,][]{1998RSPSA.454..903H,2008RvGeo..46.2006H} technique, the VMD method theoretically mitigates the influence of mode mixing issue by decomposing the signal into a sum of IMFs with analytically calculated limited center frequency and bandwidth. Mode mixing is one of the major drawbacks of EMD, which is defined as a single IMF either consisting of signals of widely disparate scales or a signal of a similar scale residing in different IMF components \citep{2008RvGeo..46.2006H}. In the VMD algorithm, the set of IMFs of an input signal, $f(t)$, is initialized as $\{u_k\}=\{u_1,...,u_K\}$ along with their corresponding center frequencies $\{\omega_k\}=\{\omega_1,...,\omega_K\}$. The objective of the VMD is to minimize the sum of bandwidths of IMFs by solving the following optimization problem\footnote{In Equation (\ref{eq:1}), $||h(x)||_2$ means the $L^2$ (Euclidean) norm of the function $h(x)$, which is calculated as $\left[\int{|h(x)|^2dx}\right]^{1/2}$. For a discrete vector $h_k(x_k)$, the $L^2$ norm of it is calculated as $\left[\sum_{k=1}^K |h_k|^2\right]^{1/2}$. From another perspective, the inner product can be expressed as $\langle p(x), q(x)\rangle=\int{p^*(x)q(x)dt}$, which implies that $||h(x)||_2=\left[\langle h(x), h(x)\rangle\right]^{1/2}$.}:
\begin{equation}
\min_{\{u_k\},\{\omega_k\}}\left\{\alpha\sum_{k=1}^K {\rm BW}_k^2 + \left|\left|f(t)-\sum_{k=1}^K u_k(t)\right|\right|_2^2\right\},\label{eq:1}
\end{equation}
where $K$ is the total number of modes, BW$_k$ is the bandwidth of the $k$th mode. The first component of the objective function enhances mode compactness, whereas the second component quantifies the reconstruction error, which ought to be minimized. The parameter $\alpha$ acts as a balancing weight between compactness and the reconstruction error. In Appendix~\ref{appendix2}, we offer a concise overview of the VMD algorithm. For the complete algorithmic description of the VMD, we recommend referring to \citet{6655981}. There are two parameters should be initially set in the VMD algorithm, which is the total number of modes $K$ and the weighting factor $\alpha$. In the VMD algorithm, selecting the parameters $K$ and $\alpha$ is vital for its efficacy. Generally, an increase in $K$ correlates with a rise in $\alpha$, suggesting that the bandwidth narrows as the number of modes grows. Following the procedure described in \citet{2023ApJ...957...84S}, we initially assign $\alpha$ a typical value of 2000 and opt for 2 modes ($K=2$). We then generate the power spectrum for each mode and compare it with the power spectrum of the original light curve. Should no mode display a power spectrum mirroring the QPO distinctive pattern, we increment $K$. This iterative strategy persist until the fundamental and harmonic components of the QPO are discernibly isolated in individual modes. Finally, we adjust $\alpha$ to align the IMF PDS profile with the QPO peak in the power spectrum of the original light curve. The selection of $K$ and $\alpha$ involves elements of subjectivity, which is one of the major concerns of the VMD algorithm \citep{7997854}. This issue requires future investigations, which are beyond the scope of this study. Nevertheless, we find that making moderate adjustments to the values of $K$ and $\alpha$ have a minimal impact on the subsequent Hilbert analysis when ensuring that the PDS of the desired IMF aligns with the QPO peak in the observed PDS (see Figure~\ref{fig:2}). In addition, the current VMD algorithm addresses certain limitations of the original EMD algorithm, and show attractive performance compared to the existing mode decomposition methods.
The code we used to perform VMD is from vmdpy v0.2\footnote{\url{https://github.com/vrcarva/vmdpy}} \citep{CARVALHO2020102073}, which is an open-source Python package. It is worth noting that an appropriate time resolution is required for the HHT analysis, as too dense sampling results in poor SNR in each bin, while too sparse sampling does not sufficiently decompose the QPO component. In our case, where the QPO frequency is $\sim$2--3 Hz and the net count rate is $\sim600$ cts s$^{-1}$, a time resolution of 0.05s is appropriate. This ensures that the time internal between the samplings is much shorter than the variability timescale, and each time bin has a signal-to-noise ratio (SNR) $\gtrsim5\sigma$. Figure~\ref{fig:2}a shows an example of a 14-s-long light curve with a 0.05s time resolution from HE observations of MAXI J1535--571 (obsID P011453500201) and the corresponding IMFs. In this representative observation, the observed light curve exhibits an LFQPO around 2 Hz. For the VMD algorithm, we selected parameters with $K=5$ and $\alpha=300$. We successfully identify a $\sim$2 Hz oscillation and its harmonic component as the second and third IMFs, respectively. To compare results of the VMD algorithm with the Fourier analysis, Figure~\ref{fig:2}b plots the power spectra for the original HE light curve alongside that of the second and third IMFs. These power spectra are generated in the frequency range of 0.04--10 Hz. A simulation also demonstrates the robustness of the VMD algorithm that it shows good performance in the presence of noise in the analyzed observations (see Appendix~\ref{appendix3} for details).

\begin{figure}
\centering
    \includegraphics[width=\linewidth]{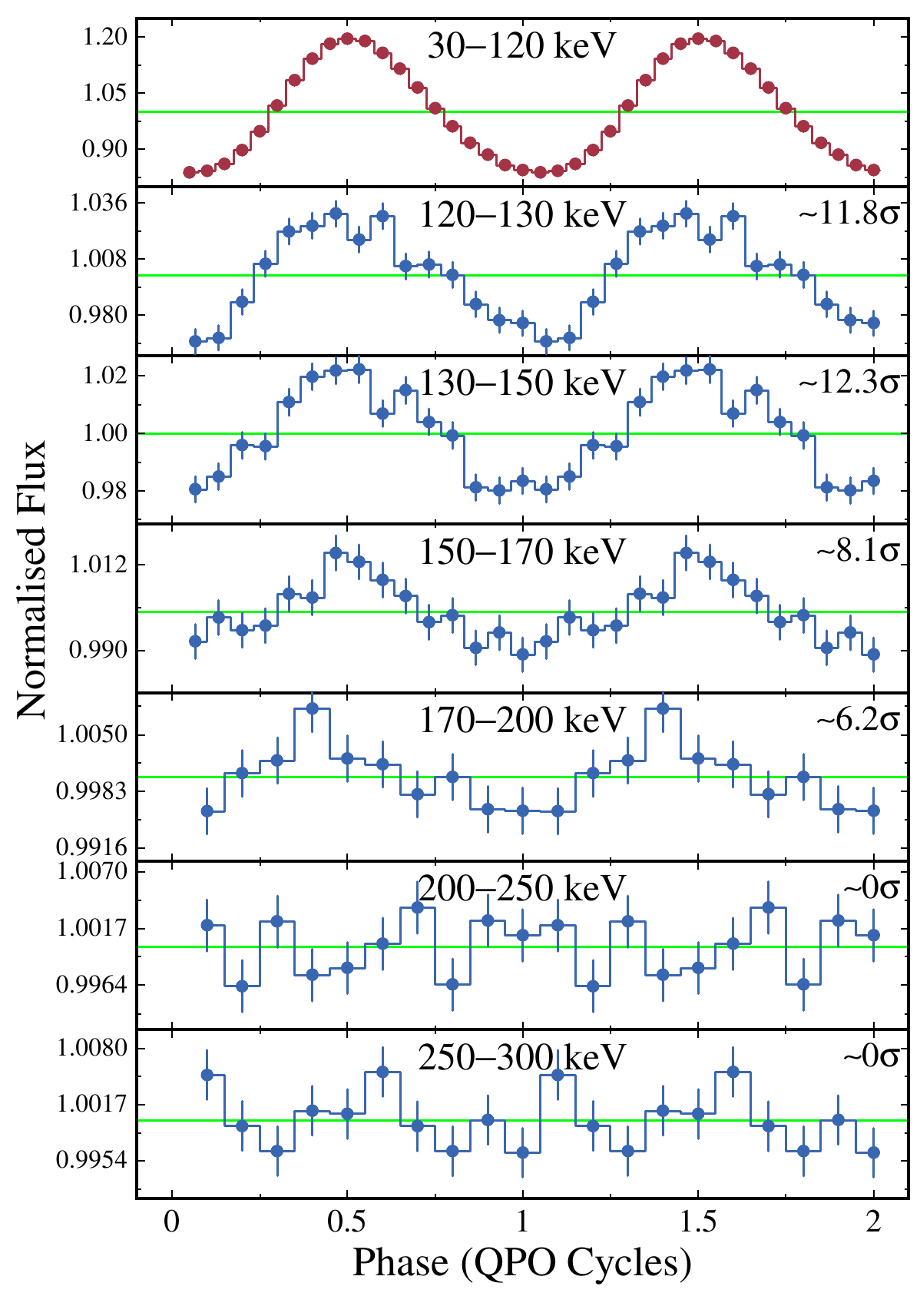}
    \caption{The Hilbert-Huang transform (HHT) phase-folded QPO waveforms of MAXI J1535--571 from Insight-HXMT/HE light curves. To enhance the SNR, waveforms from all selected observations have been collectively combined, and the final combined waveform are normalized by the mean value of the count rate. These QPO phases are obtained through HHT analysis of light curves in the 30--120 keV energy band. Each panel showcases the significance level determined via the cross-correlation technique, as detailed in Appendix~\ref{appenix}.}
    \label{fig:3}
\end{figure}

For each IMF obtained from the mode decomposition, the Hilbert transform can calculates its physically meaningful instantaneous phase, amplitude and frequency functions.
For a time series $f(t)$, its Hilbert transform is defined as 
\begin{equation}
\mathcal{H}\left[f(t)\right]=\frac{1}{\pi}{\rm p.v.}\int{\frac{f(\tau)}{t-\tau}d\tau},
\end{equation}
where $\rm p.v.$ is the Cauchy principal value. Using the Hilbert transform, one can obtain the corresponding analytic signal, which is defined as 
\begin{equation}
    f_{\rm A}(t) = f(t) + j\mathcal{H}\left[f(t)\right]=A(t)e^{j\phi(t)}, \label{eq:ana}
\end{equation}
where $j$ is the imaginary unit. The time-dependent amplitude, $A(t)$, and phase, $\phi(t)$, can be obtained from
\begin{equation}
A(t)=\left\{f^2(t)+\mathcal{H}^2\left[f(t)\right]\right\}^{1/2}    
\end{equation}
and
\begin{equation}\label{eq:phi}
\phi(t)=\arctan{\left\{\frac{\mathcal{H}(\left[f(t)\right]}{f(t)}\right\}},
\end{equation}
respectively. Therefore, the instantaneous frequency, $\nu(t)$, can be defined as
\begin{equation}
\nu(t)=\frac{1}{2\pi}\frac{d\phi(t)}{dt}.
\end{equation}
Considering the focus of the present study on investigating QPOs, we only apply the Hilbert transform to the IMF corresponding to the QPO component (e.g. the second IMF plotted in Figure~\ref{fig:2}a). From the top to bottom panels of Figure~\ref{fig:2}c, we illustrate the QPO IMF and corresponding phase function, instantaneous frequency and amplitude of the QPO component, respectively. As one can see, the instantaneous frequency of QPO fluctuates between $\sim$1 and $\sim$3 Hz over time.

\begin{figure}
\centering
    \includegraphics[width=\linewidth]{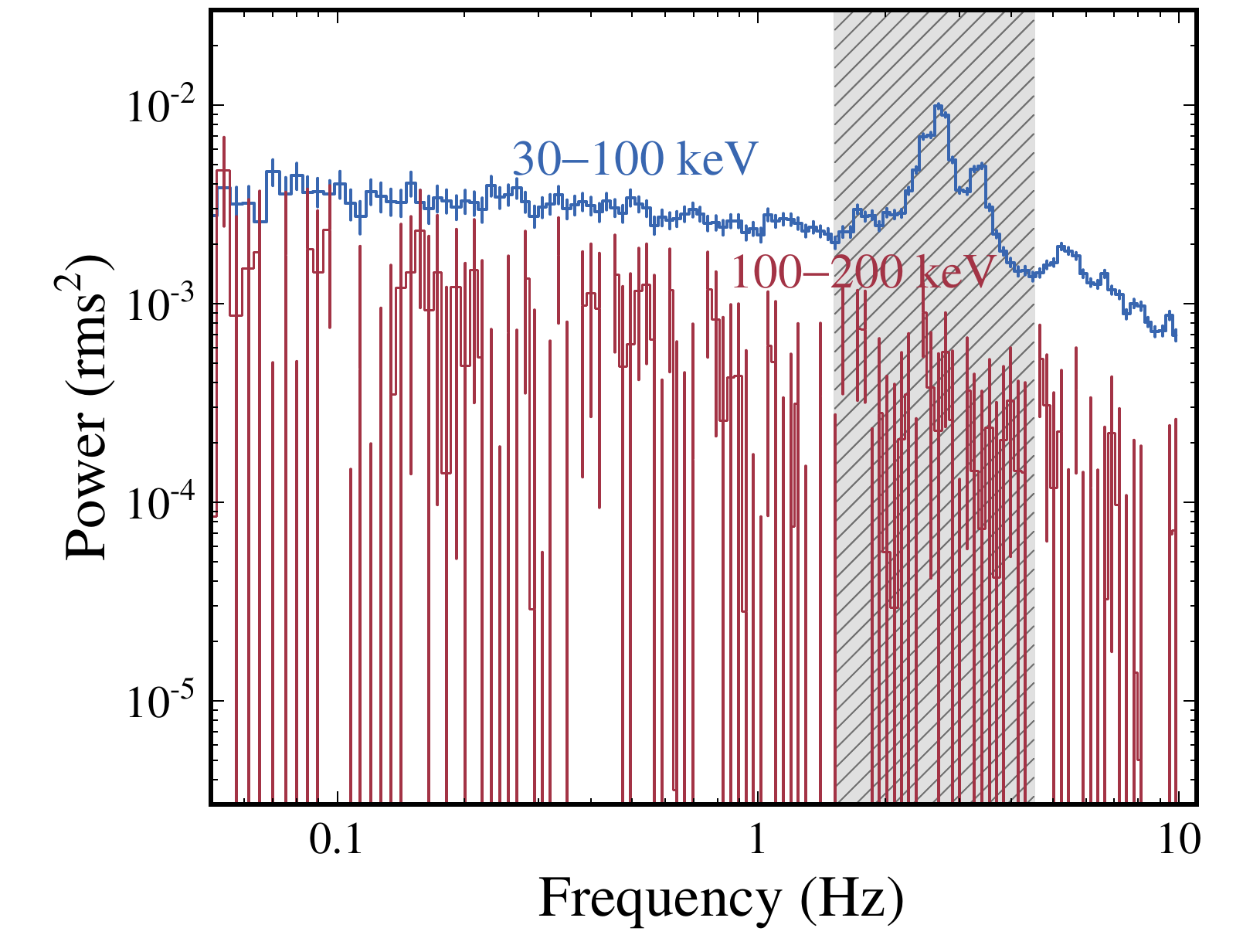}
    \caption{The long-term combined power density spectra (PDS) of 30--100 (in blue) and 100--200 (in red) keV energy bands. The shaded region highlights the frequency range where the QPO fundamental is observed.}
    \label{fig:4}
\end{figure}
\subsection{Recovery of QPO Waveforms above 120 keV}
We perform the HHT analysis for each selected observations from HE/NaI in the energy range of 30--120 keV. In such an energy band, HE shows a good performance to provide high SNR light curves. These high SNR light curves are suitable to be performed with the HHT analysis. Once the instantaneous phases, $\phi(t)$, of the QPOs have been computed using the Hilbert transform as plotted in Figure~\ref{fig:2}c, the QPO waveforms can be constructed by phase-folding the light curves \citep[also see][]{2014ApJ...788...31H,2020ApJ...900..116H,2023ApJ...957...84S}. We plot the reconstructed 30--120 keV QPO waveform in the top panel of Figure~\ref{fig:3}. To enhance the SNR, waveforms from all selected observations have been collectively combined, and the final combined waveform are normalized by the mean value of the count rate. It is evident that the QPO waveform exhibits an non-sinusoidal nature, which is characterized by a fast rise of $\sim0.4$ cycles, followed by a gradual drop of $\sim0.6$ cycles. 

Utilizing the well-defined QPO phases derived from the HHT analysis, we extend the analysis to perform phase-folding on high-energy light curves above 120 keV. It should be noted that, due to the low SNR in the high-energy band, conducting the HHT analysis there is not feasible. Consequently, the phase-folding of the high-energy light curves is carried out using the instantaneous phase obtained from the HHT analysis in the low-energy band, assuming that the high energy QPOs have the same frequency as that in the low energy band. From the second to the bottom panels of Figure~\ref{fig:3}, the phase-folded light curves in various energy bands above 120 keV are plotted. We use two methods to determine the significance of the high energy QPO profiles, one is the standard $\chi^2$-test method, another is the cross-correlation technique described in \citet{2022ApJ...938..149H}. It is evident that the QPO profile remains distinctly identifiable in the energy bands between 120 and 200 keV. However, signals dissipate in energy ranges exceeding 200 keV. The significance of the QPO profile in the 150--170 keV energy band derived from the $\chi^2$-test method is $5.2\sigma$, while the cross-correlation technique gives a $8.1\sigma$ confidence level. As for the QPO profile in 170--200 keV, the cross-correlation technique gives a $6.2\sigma$ confidence level (see Appendix~\ref{appenix} for details). However, the $\chi^2$-test method gives only $2.1\sigma$ for the 170--200 keV profile. This indicates that the cross-correlation method is more advantageous since it takes the lower-energy QPO profile, which has a very high SNR, as a template in searching for the higher-energy pulsed signals \citep[see also][]{2022ApJ...938..149H}. We also plot the PDS for high-energy light curves (100--200 keV) to provide a comparison between the HHT phase-folding and Fourier methods. Notably, no significant QPO peaks are observed in the long-term combined PDS above 100 keV, as shown in Figure~\ref{fig:4}. In a simulation study presented in Appendix~\ref{appendix3}, we find that a small part of the noise component is decomposed into the QPO IMF, resulting from mode mixing. However, the simulation results show that if the QPO signals do not exist in the 170-200 keV energy band, the significance of the modulation from the noise is found to be $<2\sigma$, which is inconsistent with the observational results ($\sim 6.2\sigma$). Conversely, when QPO signals are added to the high-energy light curves, the significance of the modulation in the phase-folded light curve is found to be $\sim6-8\sigma$, consistent with the observational results. This provides supports for the idea that the observed modulations of the phase-folding profile in the energy ranges of 170--200 keV could indeed originate from the QPO component, despite the presence of some mode mixing issues in the VMD. (see Appendix~\ref{appendix3} for details).

\subsection{Properties of High Energy QPOs}
\begin{figure*}
\centering
    \includegraphics[width=0.8\textwidth]{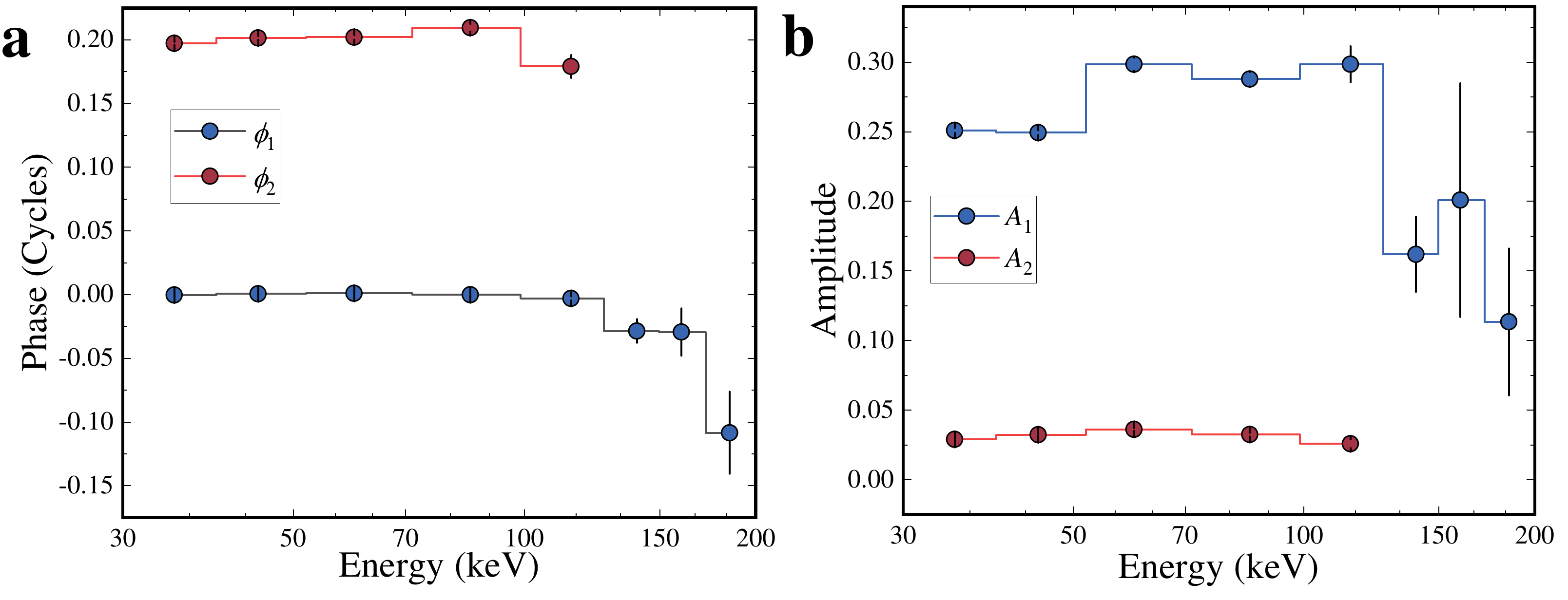}
    \caption{The best-fitting phases (a) and amplitudes (b) of two harmonically related cosine functions are plotted as a function of the photon energy.} \label{fig:5}
\end{figure*}

It is crucial to highlight that the 170--200 keV QPO signal, as reported, marks the highest-energy QPO detection for this source to date. To further study the properties of the high-energy ($>$120 keV) QPO profiles, we analyze the phase-folded waveforms using harmonically related cosine functions for fitting. From Figure~\ref{fig:3}, it is evident that the QPO waveforms exhibit a non-sinusoidal nature, particularly below 130 keV, which is consistent with the emergence of the second harmonic in the PDS (see Figure~\ref{fig:2}b). We find that two cosine functions provide a good description of the data. Furthermore, we subtract the background contribution from the fractional amplitude, resulting in the final fitting function of
\begin{equation}
f(\phi)=1-\frac{S}{S+B}\{A_1\cos{(\phi-\phi_1)}+A_2\cos{\left[2(\phi-\phi_2)\right]}\},    
\end{equation}
where $A_1$ and $A_2$ represent the amplitudes, $\phi_1$ and $\phi_2$ the phases of the cosine functions, and $S$ and $B$ denote the mean count rates of the source and background, respectively. We perform phase-folding and subsequent model fitting in 8 distinct energy bands of 30--40, 40--50, 50--70, 70--100, 100--130, 130--150, 150--170, and 170--200 keV. The analysis excludes waveforms above 200 keV due to the insignificance of the QPO signal in these bands. Notably, the parameters of the second harmonic component ($A_2$ and $\phi_2$) are effectively constrained only at energies below 130 keV. This suggests a notable attenuation of the second harmonic component above 130 keV, prompting us to employ one cosine function for fitting in these energy bands. Figure~\ref{fig:5} presents the energy dependence of the best-fitting parameters. It is clear that $\phi_1$ and $\phi_2$ remains relatively stable from 30 to 130 keV, whereas $\phi_1$ exhibits a decreasing trend with increasing energy beyond 130 keV, indicating soft lags across different energy bands. Regarding the amplitudes, while $A_2$ stays relatively constant, $A_1$ shows a slight increase from 30 to 130 keV, followed by a decrease from $\sim0.3$ to $\sim0.1$ above 130 keV.


\section{Discussion} \label{sec:4}
In this study, we have performed a comprehensive analysis of type-C QPOs in the black hole binary MAXI J1535--571, utilizing data from the Insight-HXMT. Based on the Hilbert-Huang transform (HHT) method, we have proposed a novel method for detecting high-energy QPO signal. This allows us to successfully reconstruct the QPO waveform above 100 keV through phase-folding of the light curves, although no significant QPO signal has been discernible in the PDS beyond this threshold. Notably, by combining QPO waveforms derived from seven observations, we have achieved a highly confident detection of QPO profiles above 170 keV at a significance level of $\sim6.2\sigma$. This represents the highest-energy detection of QPOs in this source reported to date. Furthermore, this technique has the potential to extend the study of energy-dependent behaviors of QPO amplitude and phase lag up to 200 keV. 

The general method for detecting QPO signals from X-ray light curves is the Fourier transform, which allows QPOs to be identified as a series of harmonic peaks in the PDS. However, in the high-energy bands ($>$100 keV), the limitations in count statistics present a challenge. Previous studies commonly combined multiple observations to enhance the SNR of QPO peaks in the PDS \citep[see e.g.][]{2021NatAs...5...94M,2022ApJ...932....7Y}. However, this approach assumes similar QPO frequencies across the combined observations, which is not typically the case during the HIMS of outburst rising phase of BHXRBs, where the QPO frequency tends to increase rapidly \citep[see e.g.][]{2018ApJ...866..122H,2021MNRAS.508..287S,2023ApJ...943..165S}. If the QPO frequencies among combined observations show significant differences, multiple fundamental QPO peaks would appear in the long-term combined PDS in the low-energy band (see Figure~\ref{fig:4}). In such cases, improving the SNR of higher energy QPO signals would be challenging, as the QPO signals from different observations are distributed over a wide frequency range. Although any method would face with the limitations in count statistics in the high-energy bands, the HHT phase-folding method proposed in this work effectively addresses the aforementioned issue encountered in the Fourier method when combining different observations. In the phase-folding approach, the well-defined QPO phases can be obtained in the low-energy band. This indicates that, for any QPO frequency, a QPO cycle can be divided into given phase bins between 0 and $2\pi$. Consequently, by stacking up counts within the same phase bin from multiple observations, one can achieve a higher SNR associated with the modulation of the count rate as function of the phase (i.e. the QPO waveform). Both our observational and simulation studies demonstrate that, provided the phase information of the QPO, the phase-folding approach performs better in determining the high-energy QPO signal than the Fourier method when the QPO frequency varies among different observations (see Appendix~\ref{appendix3}).

Previous studies by \citet{2018ApJ...866..122H} and \citet{2020JHEAp..25...29K} have presented the energy dependence of type-C QPO amplitude in MAXI J1535--571 across a wide energy band ranging from 1 to 100 keV. They observed an increase in QPO RMS amplitude with photon energy up to $\sim$20 keV, after which it plateaus across all observations. Utilizing our HHT-based technique, we extend this study to analyze QPO amplitude beyond 100 keV. As shown in Figure~\ref{fig:5}b, we note a marginal increase in amplitude from 30 to 60 keV, followed by a plateau between 60 and 120 keV, and a decreasing trend beyond 120 keV. The flat trend of amplitude between 30 and 120 keV agrees with findings reported by \citet{2018ApJ...866..122H} and \citet{2020JHEAp..25...29K}.

In broad-band spectral analyses of MAXI J1535--571, both \citet{2020JHEAp..25...29K} and \citet{2022ApJ...935...25R}, using Insight-HXMT and INTEGRAL/SPI data respectively, reported findings indicative of a relatively low-temperature corona. \citet{2020JHEAp..25...29K} reported the cutoff energy of the power-law component within the 60–80 keV range, while \citet{2022ApJ...935...25R} found the temperature of thermal electrons in the corona to be $\sim20$ keV during the HIMS. These observations suggest that emissions above $\sim$100 keV are unlikely to be corona-dominated. Furthermore, using INTEGRAL/SPI data, \citet{2022ApJ...935...25R} identified an additional high-energy component, which can be modeled by a cutoff power-law and is observed to have a flux approximately six times over the coronal flux in the 100–400 keV range (refer to Figure 6 in \citet{2022ApJ...935...25R}). Consequently, it is probable that the QPO amplitude above 100 keV is predominantly influenced by this extra high-energy component.
The different evolutionary trends between the coronal and extra power-law fluxes, as reported by \citet{2022ApJ...935...25R}, imply different origins for these two components, with the high-energy component likely being from a jet. This is further corroborated by the divergent energy-dependent properties of both $\phi_1$ and $A_1$ in low- (30–130 keV) and high-energy (130–200 keV) bands, as shown in Figure~\ref{fig:5}. Particularly, the emergence of soft lags above 130 keV agrees with predictions made by the jet precession model \citep{2021NatAs...5...94M,2023ApJ...948..116M,2023ApJ...957...84S}. 

Using the phase function in 30--120 keV to fold the light curves above 120 keV appears to introduce some systematic issues, as spectral analyses have indicated different dominant components between the two energy bands. However, our results demonstrate that the high-energy QPO waveform can be effectively reconstructed using the 30--120 keV phase function, suggesting correlated QPO phases across these energy bands. We propose that a jet-corona coupling precession scenario could explain our findings, wherein the jet and corona are coupled and precess with the same period, leading to correlated QPO signals from the two locations. This picture is further supported by a recent general relativistic magneto-hydrodynamic simulation \citep{2018MNRAS.474L..81L}.

\begin{figure*}
\centering
    \begin{minipage}[c]{0.32\textwidth}
\centering
    \includegraphics[width=\linewidth]{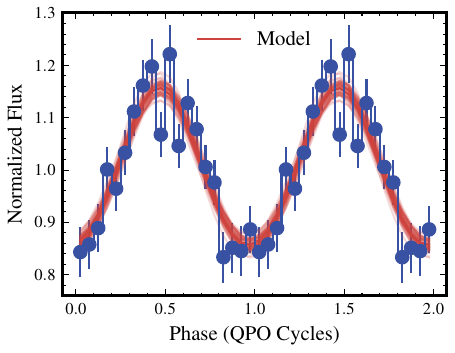}
\end{minipage}
\begin{minipage}[c]{0.32\textwidth}
\centering
    \includegraphics[width=\linewidth]{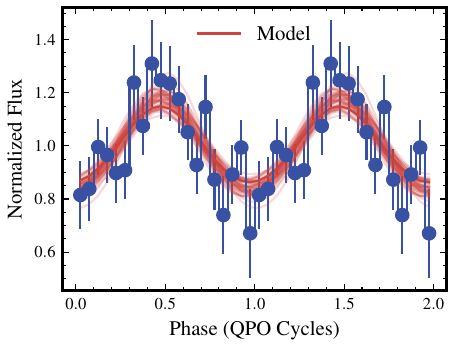}
\end{minipage}
\centering
    \begin{minipage}[c]{0.32\textwidth}
\centering
    \includegraphics[width=\linewidth]{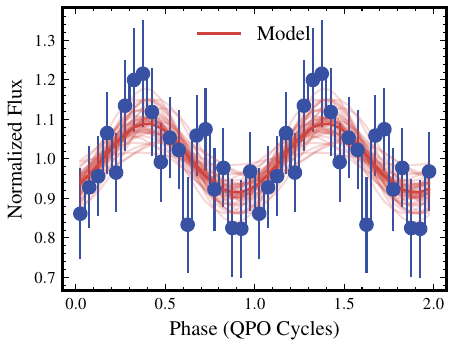}
\end{minipage}
    \caption{The fitting results of the waveforms with the jet precession model in the energy bands between 130 and 200 keV. Sequentially from left to right, the fitting results for the energy bands of 130--150 keV, 150--170 keV, and 170--200 keV are depicted. All data points displayed have undergone background subtraction. The red solid lines represent the jet precession model plotted from the posterior distribution derived with the MCMC method.}
    \label{fig:6}
\end{figure*}

Given that typical speeds at the base of the jet can already be relativistic \citep{2006MNRAS.368.1561M}, a fundamental concept in the jet precession model is that emission along the jet axis is considerably stronger than off-axis emission due to the Doppler boosting effect \citep[see][]{2021NatAs...5...94M}. As a result, when the jet precesses, the flux from it undergoes modulation. The velocity of the jet material significantly influences the QPO variability. In particular, higher jet speeds result in greater modulation amplitude, as there is a larger difference in emission strength between on-axis and off-axis situations. During the stage of the studied observations, it was reported to exist a compact and continuous jet in MAXI J1535--571 from the radio band \citep{2019ApJ...883..198R}. For a jet of this nature, the observed luminosity, $S_{\rm obs}$, is described by
\begin{equation}
S_{\rm obs} = S_{\rm int}\mathcal{D}^{\Gamma+2},
\end{equation}
where $S_{\rm obs}$ represents the observed luminosity, $S_{\rm int}$ is the intrinsic luminosity of the jet, $\Gamma$ is the spectral index of the power-law spectrum of the jet emission, and $\mathcal{D}$ is the relativistic boosting factor given by
\begin{equation}
\mathcal{D}=\frac{(1-\beta^2)^{1/2}}{1-\beta\cos{\theta}},
\end{equation}
where $\beta=v/c$ is the jet speed in units of the speed of light, and $\theta$ is the angle between the jet axis and the observer's line of sight, calculated as
\begin{align}
\cos{\theta} = &\sin{i}\cos{\Phi}\sin{\gamma}\cos{\omega}+\sin{i}\sin{\Phi}\sin{\gamma}\sin{\omega}\notag\\
&+ \cos{\gamma}\cos{i},
\end{align}
where $i$ is the inclination angle between the observer's line of sight and the black hole spin axis, $\Phi$ is the azimuth angle of the observer, $\gamma$ is the angle between the jet axis and the black hole spin axis, and $\omega$ is the precession phase angle, ranging from 0 to 2$\pi$ over the precession period. \citep[see][for details]{2021NatAs...5...94M}. Setting the inclination, $i$, to $57^\circ$ \citep[][]{2018ApJ...852L..34X}, and assuming $\gamma$ is $5^\circ$ \citep[see also][]{2021NatAs...5...94M,2023ApJ...957...84S} and $\Gamma=1.6$ which is the spectral index of the hard tail component \citep{2022ApJ...935...25R}, we employ the Markov Chain Monte Carlo (MCMC) method with a 50,000 sample length to fit the QPO waveforms using the jet precession model. The model has two free parameters: $\beta$ and $\Phi$. Given that the QPO radiative mechanism above 130 keV is predominantly jet-driven, while below 130 keV it could be affected by coronal emission, our fitting is only performed on the 130--150, 150--170, and 170--200 keV energy ranges. The fitting results are presented in Figure~\ref{fig:6}. Based on the fitting with the jet precession model, we find $\beta=0.44\pm0.02$, $\beta=0.47\pm0.06$ and $\beta=0.30\pm0.07$ for 130--150, 150--170 and 170--200 keV energy bands, respectively. The value of $\Phi$ depends on the reference point of the precession cycle, and the fitting results of $\Phi=170^\circ.0\pm3^\circ.7$, $\Phi=171^\circ.9\pm8^\circ.9$ and $\Phi=143^\circ.4\pm15^\circ.2$ for 130--150, 150--170 and 170--200 keV energy bands, respectively. The phase differences between these energy bands represent soft lags. Based on the radio observations, \citet{2019ApJ...883..198R} suggested an inclination of $\leqslant45^{\circ}$. We also perform fittings with the jet precession model, assuming $i=45^{\circ}$. The best-fitting parameters for $\beta$ are $0.46\pm0.02$, $0.49\pm0.06$, and $0.32\pm0.07$ for the 130--150, 150--170, and 170--200 keV energy bands, respectively, closely resembling the values obtained when $i=57^\circ$. The uncertainties are reported at the $1\sigma$ confidence level.

\section{Summary and Conclusion}\label{sec:5}
A detailed analysis of type-C QPOs from the black hole binary MAXI J1535--571 with observations from Insight-HXMT is presented in this work. We propose a novel method based on the Hilbert-Huang transform, which enables a robust detection of type-C QPOs up to 200 keV, and consistently above 170 keV. This method also has the potential to extend the study of the energy dependence of QPO amplitude and phase lag up to 200 keV. Our findings reveal that the QPO fundamental exhibits soft lags and a decrease in amplitude above $\sim100$ keV. Given the reported detection of a hard tail in broad spectroscopy, we proposed that the QPO properties above 100 keV are predominantly influenced by the hard tail component, likely originating from a relativistic jet. Since the high-energy QPO waveform can be well constructed using the phase function derived from the low-energy light curves, a jet-corona coupling precession scenario could explain our findings. Fitting the phase-folded profile with jet precession model suggested a jet velocity of $\sim0.4c$. 
\begin{acknowledgments}
We are grateful to the anonymous referee for constructive comments that helped us improve this paper.
This research has made use of data obtained from the High Energy Astrophysics Science Archive Research Center (HEASARC), provided by NASA’s Goddard Space Flight Center, and the Insight-HXMT mission, a project funded by China National Space Administration (CNSA) and the Chinese Academy of Sciences (CAS). This work is supported by the National Key R\&D Program of China (2021YFA0718500) and the National Natural Science Foundation of China under grants, U1838201, U1838202, 12173103, U2038101 and U1938103. This work is partially supported by International Partnership Program of Chinese Academy of Sciences (Grant No.113111KYSB20190020). Ling D. Kong is grateful for the financial support provided bu the Sino-German (CSC-DAAD) Postdoc Scholarship Program (57251553).
\end{acknowledgments}

%




\appendix
\section{VMD technique}
\label{appendix2}
VMD is a recently proposed technique designed to decompose a time series into several intrinsic mode functions \citep{6655981}. In the HHT method, the original mode decomposition technique is the EMD \citep{1998RSPSA.454..903H}, which is an algorithm based on interpolation of local extrema. However, this algorithm has a significant drawback: the problem of mode mixing, where widely different scales could appear in a single IMF component, or a coherent signal fragmented into separate parts that appear in more than one IMF component. In comparison with the traditional EMD method, the VMD algorithm theoretically reduces the impact of mode mixing by simultaneously decomposing the input signal into a sum of IMFs while analytically determining the limited center frequency and bandwidth.

In the VMD, the IMFs are represented as amplitude-modulated-frequency-modulated (AM-FM) signals, given by $u_k(t)=A_k(t)\cos{(\phi_k(t))}$, where the phase for the $k$th mode, $\phi_k(t)$, is a non-decreasing function, and the envelope $A_k(t)$ is non-negative ($A_k(t)\geq0$). When applied to an input signal $f(t)$, the VMD initializes the set of all modes as $\{u_k\}=\{u_1,...,u_K\}$ along with their corresponding center frequencies $\{\omega_k\}=\{\omega_1,...,\omega_K\}$. The goal is to minimize the sum of bandwidths of IMFs by solving the optimization problem (\ref{eq:1}). To estimate the bandwidth of each mode (BK$_k$), the VMD employs  several signal processing tools. Details on these signal processing tools can be found in \citet{6655981}. The bandwidth of each mode is estimated using $H^1$ Gaussian smoothness of the shifted signal, i.e. the $L^2$ norm of the gradient:
\begin{equation}
{\rm BW}_k = \left|\left|\partial_t\left[\left(\delta(t)+\frac{j}{\pi t}\right)*u_k(t)\right]e^{-j\omega_kt}\right|\right|_2,
\end{equation}
where $\delta$ is the Dirac function, and `$*$' denotes convolution. The term enclosed in the square brackets represents the analytic signal of $u_k(t)$. With Lagrangian multipliers, $\lambda$, the augmented Lagrangian, $\mathcal{L}$, can be expressed as
\begin{align}
\mathcal{L}(\{u_k\}, \{\omega_k\}, \lambda)=\alpha\sum_{k=1}^K {\rm BW}_k^2 + \left|\left|f(t)-\sum_{k=1}^K u_k(t)\right|\right|_2^2+\left\langle \lambda(t), f(t)-\sum_{k=1}^K u_k(t)\right\rangle.
\end{align}
The solution to the original minimization problem (\ref{eq:1}) is now achieved as the saddle point of the augmented Lagrangian, which can be found using the alternate direction multiplier method (ADMM). Then all the modes gained from solutions in Fourier domain are written as
\begin{equation}
    \hat{u}_k(\omega)=\frac{\hat{f}(\omega)-\sum_{i\neq k}{\hat{u_i}(\omega)} + \frac{\hat{\lambda}(\omega)}{2}}{1+2\alpha(\omega-\omega_k)^2},
\end{equation}
where $\hat{f}(\omega)$ and $\hat{u}_k(\omega)$ represent the Fourier transform of the signal $f(t)$ and mode $u_k(t)$, respectively, $\omega$ is the Fourier frequency. The VMD algorithm iteratively refines the modes $\{u_k\}$ and their corresponding center frequencies $\{\omega_k\}$ until they stabilize. In this algorithm, the center frequency $\omega_k$ is determined within the Fourier domain by the formula:
\begin{equation}
    \omega_k = \frac{\int_0^\infty\omega \left|\hat{u}_k(\omega)\right|^2d\omega}{\int_0^\infty\left|\hat{u}_k(\omega)\right|^2d\omega}.
\end{equation}
This approach effectively extends Wiener filtering to accommodate adaptive filtering across multiple frequency bands. For a comprehensive exploration of this technique and the constrained variational optimization problem, see \citet{6655981}.

\section{Simulation Analysis}
\label{appendix3}

\begin{figure*}
\centering
\includegraphics[width=0.45\textwidth]{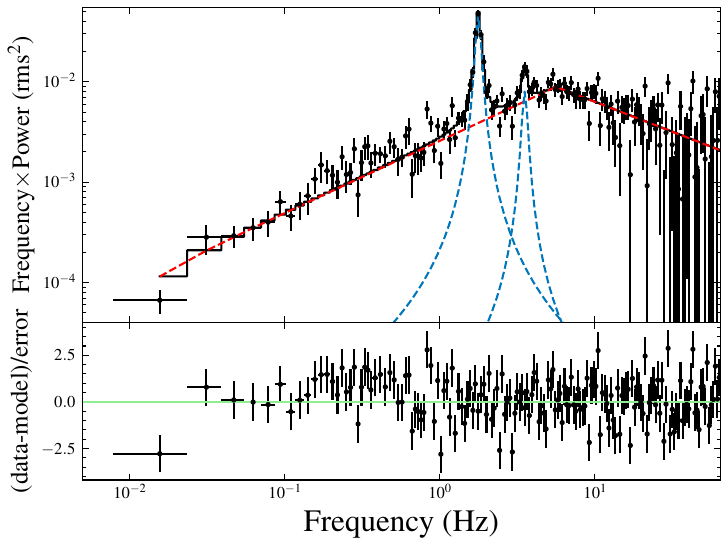}
    \caption{Power density spectrum of the HXMT/HE observation P011453500201 in the 30--120 keV energy band fitted with a model consists of a broken power law function (plotted in red) and two Lorentzian functions (plotted in blue). In the plotted power density spectrum, Poisson noise has been subtracted.} \label{fig:7}
\end{figure*}

\begin{figure*}
\centering
    \begin{minipage}[c]{0.45\textwidth}
\centering
    \includegraphics[width=\linewidth]{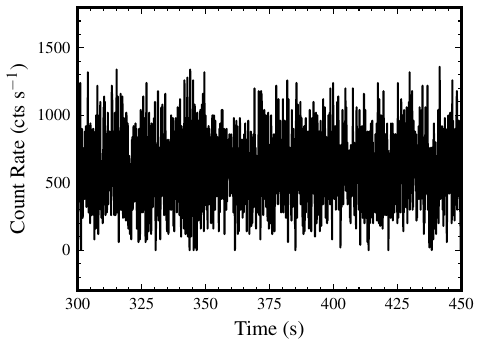}
\end{minipage}
\begin{minipage}[c]{0.45\textwidth}
\centering
    \includegraphics[width=\linewidth]{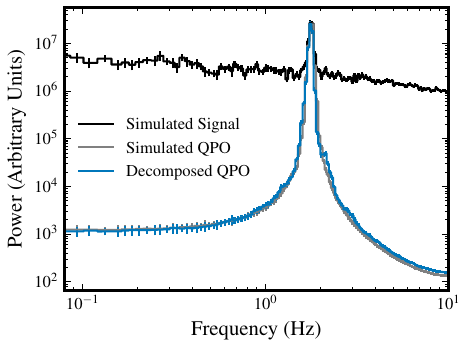}
\end{minipage}
    \caption{\textit{Left}: simulated light curve composed of a red noise and a QPO component. In this simulated signal, the Poisson noise is also considered. \textit{Right}: Fourier power spectra that are produced from the original total light curve (in black), simulated QPO component (in grey), and decomposed QPO component (in blue), respectively.} \label{fig:B1}
\end{figure*}

\begin{figure*}
\centering
    \begin{minipage}[c]{0.4\textwidth}
\centering
    \includegraphics[width=\linewidth]{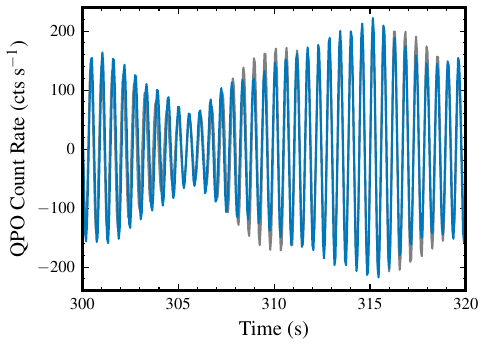}
\end{minipage}
\begin{minipage}[c]{0.4\textwidth}
\centering
    \includegraphics[width=\linewidth]{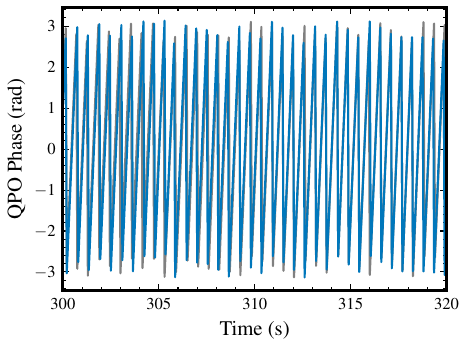}
\end{minipage}\\
\centering
    \begin{minipage}[c]{0.4\textwidth}
\centering
    \includegraphics[width=\linewidth]{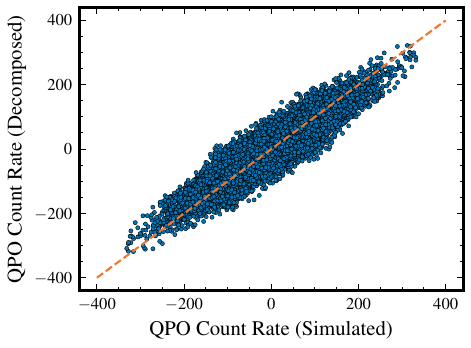}
\end{minipage}
\centering
    \begin{minipage}[c]{0.4\textwidth}
\centering
    \includegraphics[width=\linewidth]{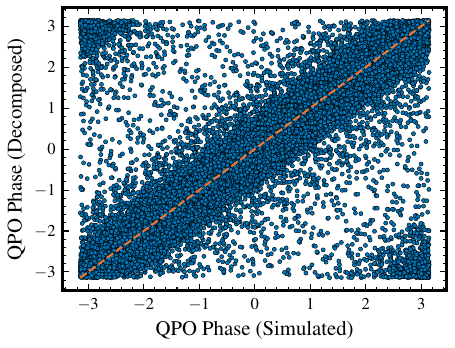}
\end{minipage}\\
    \begin{minipage}[c]{0.4\textwidth}
\centering
    \includegraphics[width=\linewidth]{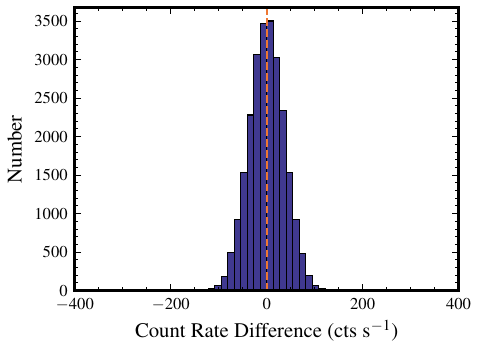}
\end{minipage}
    \begin{minipage}[c]{0.4\textwidth}
\centering
    \includegraphics[width=\linewidth]{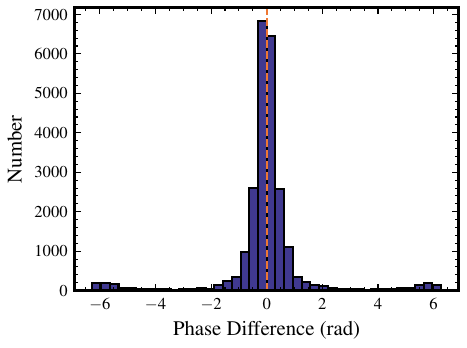}
\end{minipage}
    \caption{The top panels respectively compare the time variations of the QPO count rate and instantaneous phase between the decomposed QPO IMF (in blue) and synthetic QPO signal (in grey) in a representative 20-second time range. The two middle panels present the parameter correlations between decomposed and synthetic QPO signals using all time bins. The two bottom panels show the histograms of count rate difference and phase difference between the simulated QPO and decomposed QPO signals from all time bins. The orange dashed lines displayed in the middle and bottom panels represent the case that the two signals are in perfect alignment.} \label{fig:B2}
\end{figure*}

To assess the reliability of the QPO phase-folding analysis results, we conduct a comprehensive simulation analysis for our HHT-based method to test its robustness. This simulation analysis consists of three parts: (1) testing the performance of the VMD algorithm to verify that a simulated QPO, as strong as the observational one, is indeed decomposed accurately; (2) exploring how the VMD algorithm deals with a ``pure noise'' signal; and (3) addressing a potential issue that may arise from (1) and (2) in the phase-folding analysis.

While it has now become standard practice to fit power spectra in X-ray binaries with a combination of both broad and peaked components modeled as Lorentzians \citep[see e.g.][]{2002ApJ...572..392B,2022ApJ...932....7Y,2022MNRAS.515.1914Z,2023ApJ...943..165S}, there is currently no physical backing behind the use of Lorentzian components. Additionally, we find the continuum of the power spectrum from obsID P011453500201 in 30--120 keV can be well fitted with a broken power law function (see Figure~\ref{fig:7}). Consequently, we individually model the red noise with a broken power law function and QPO harmonics with the Lorentzian functions. Using the function shapes derived from the modeling of the observed power spectrum, we apply the algorithm from \citet{1995A&A...300..707T} to generate 1684-s light curves with a 0.05-s time resolution for both red noise and QPO components. In this algorithm, the power spectral amplitude at each frequency is randomly drawn from a $\chi^2$ distribution, and not fixed at the amplitude of the underlying power spectrum. Additionally, the Fourier phases given to the signal by this algorithm is random. For the broken power law funtion of the red noise spectrum, we set the two power law indices at 0.28 and 1.59, respectively, the break frequency at 5.61 Hz, and the fractional rms amplitude at 30\%. As for the Lorentzian function of the QPO component, we set the centroid frequency at 1.78 Hz, the FWHM at 0.12 Hz, and the fractional rms amplitude at 15\%, respectively. The simulations of the light curves are performed using Stingray\footnote{\url{https://github.com/StingraySoftware/stingray/tree/v1.1.2}}, which is an open-source Python package \citep{2019ApJ...881...39H}. In this study, we focus on the QPO fundamental and therefore do not include the harmonic component in the simulation analysis. The total light curve with an average count rate of 600 cts s$^{-1}$ consists of the red noise, QPO and Poisson noise, which is presented in the left panel of Figure~\ref{fig:B1}. Subsequently, the VMD technique is used to decompose this light curve. It is also important to note that Poisson noise is the observational errors caused by statistical fluctuations in photon counting, while red noise is related to the intrinsic stochastic behavior of the source. Therefore, Poisson noise is specifically incorporated by Poisson sampling of the variable source flux which includes both the red noise and QPO components. The right panel of Figure~\ref{fig:B1} shows the power spectra of the simulated total (in black), QPO (in grey) light curves and decomposed QPO IMF (in blue). As one can see, the power spectrum of the decomposed QPO IMF is well consistent with that of the simulated QPO signal. This demonstrates that the VMD can precisely recover the essential features of the QPO in the Fourier domain, including the centroid frequency, width and amplitude. We then perform additional Hilbert transform analyses on both the decomposed and simulated QPO light curves. This is followed by comparing the trends in QPO count rate and instantaneous phase between the two sets of light curves. The top panels of Figure~\ref{fig:B2} present a comparison of the time variations in QPO count rate and instantaneous phases between the decomposed (in blue) and simulated (in grey) QPO signals in a representative 20-s time range. The middle panels present the parameter correlations between the two signals, incorporating all time bins. In addition, the two bottom panels show the histograms of count rate difference and phase difference between the simulated QPO and decomposed QPO signals, respectively. The orange dashed lines displayed in the middle and bottom panels represent the case that the two signals are in perfect alignment. We find the structures of the decomposed QPO are highly consistent with those of the simulated QPO signal, but with local fluctuations, possibly due to mode mixing between the QPO and noise components, or the statistical fluctuations. Specifically, in the right middle panel of Figure~\ref{fig:B2}, a small number of data points are distributed in the upper left and lower right of the plot, corresponding to the phase difference distributed around $-2\pi$ and $2\pi$ in the right bottom panel. Since the phase of the QPO varies from $-\pi$ to $\pi$, this phenomenon reflects that the phases of the simulated and decomposed QPO signals in a small part of time bins have local shifts.

\begin{figure*}
\centering
\includegraphics[width=0.45\textwidth]{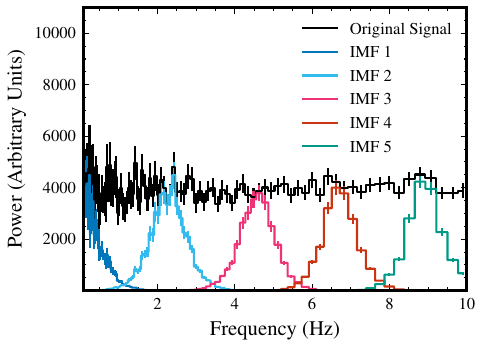}
    \caption{Power density spectra of the simulated pure Poisson noise (in black) and the corresponding VMD IMFs (in colors). The decomposed IMFs are evenly distributed in frequencies, suggesting that VMD is essentially a set of adaptive Wiener filter bank capable of separating modes with different center frequencies.} \label{fig:B3}
\end{figure*}
The VMD technique is also applied to a pure Poisson noise signal. In Figure~\ref{fig:B3}, we plot the power spectra of the Poisson noise and corresponding VMD IMFs. The result shows that the decomposed IMFs are evenly distributed in frequencies, suggesting that VMD is essentially a set of adaptive Wiener filter bank capable of separating modes with different center frequencies. The equivalent filter bank properties of the VMD have been extensively studied in literature \citep[see e.g.][]{WANG2015243}. From another perspective, the VMD algorithm is capable of generating the output IMFs for any given signal, even if the signal is pure noise. This indicates that the IMF corresponding to the QPO component might be influenced by the noises, potentially leading to local variations between the real and decomposed QPO signals observed in Figure~\ref{fig:B2}.

\begin{figure*}
\centering
\includegraphics[width=0.45\textwidth]{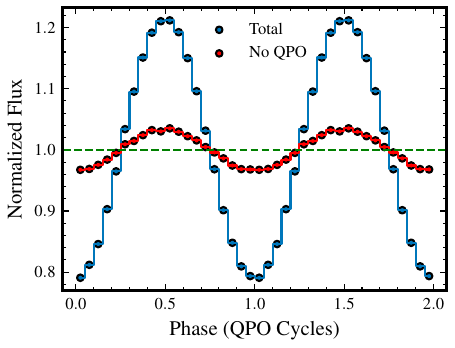}
    \caption{Phase-folding profiles of the simulated total (in blue) and noise (in red) light curves. The phase-folding is performed by using the instantaneous phase function of the decomposed QPO IMFs and by combining the seven simulated light curves (see the corresponding text for details).} \label{fig:B4}
\end{figure*}

\begin{figure*}
\centering
\includegraphics[width=0.45\textwidth]{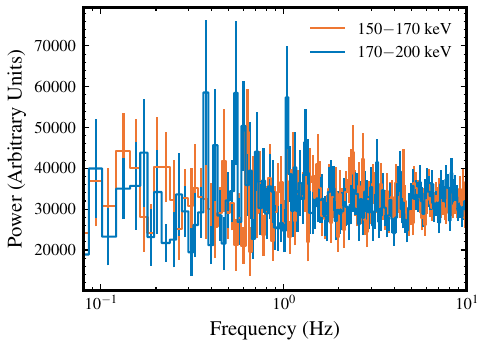}
    \caption{Power density spectra of the simulated light curves in the energy ranges of 150--170 (in orange) and 170--200 (in blue) keV.} \label{fig:B5}
\end{figure*}

To explore the potential impact of the local variations between simulated and decomposed QPO light curves presented in Figure~\ref{fig:B2} on the phase-folded light curve, we conduct the phase-folding on the initially simulated total and QPO-subtracted light curves, utilizing the instantaneous phase of the decomposed QPO IMF. Here, the QPO-subtracted light curve refers to the red noise with its Poisson noise. In our observational study, we combined data from seven different observations, each with varying QPO frequencies and exposures. Our simulation analysis is designed to closely replicate the observational analysis, taking into account these variations in QPO frequencies and exposures as accurately as possible. Therefore, we simulate seven sets of light curves in which the QPO frequency and exposure times are set to be the same as those in the corresponding observations. For each total light curve, we apply the VMD method and the Hilbert transform. Subsequently, the phase-folded total and QPO-subtracted light curves are obtained by phase-folding across the seven light curves using the QPO instantaneous phases obtained from the HHT analysis. As shown in Figure~\ref{fig:B4}, the phase-folding profile of the total light curve exhibits a significant modulation, which is defined as the QPO waveform. The amplitude of the modulation is estimated to be $\sim0.21$, with a corresponding rms of $\sim14.9$\%, consistent with the initially set value in the simulation (15\%). However, a relatively weaker modulation (with an amplitude of $\sim0.04$) of phase-folded QPO-subtracted light curve is also observed. This indicates that a small portion of the red noise is decomposed into the QPO IMF, resulting from the issue of mode mixing.

As the phase-folded noise light curve simulated in the low-energy band can exhibit modulations due to mode mixing (see Figure~\ref{fig:B4}), it is important to confirm that the observed modulations in the high-energy bands (see Figure~\ref{fig:3}) are indeed from the QPO component and not from the red noise. To achieve this, we simulate high-energy light curves for the seven observations based on the conditions of the observational data. Specifically, in the 150--170 and 170--200 keV energy bands, the average total count rates of the observed light curves are $\sim32$ and $47$ cts s$^{-1}$, respectively, with the net count rates of $\sim2$ cts s$^{-1}$ in both energy bands. For simplicity, we assume that the light curves in different energy bands for both the QPO and red noise are fully coherent. This indicates that the light-curves of both the QPO and red noise have consistent phase functions across different energy bands. Additionally, we assume that the fractional rms of the red noise and the QPO remains constant with photon energy. Therefore, to simulate red noise and QPO light curves in the two high-energy bands, we scale the absolute amplitude of the original simulated red noise and QPO light curves in the 30--120 keV energy band by a factor of 1/300 ($\sim2$ cts s$^{-1}$/$\sim600$ cts s$^{-1}$). Additionally, in these simulated total light curves, Poisson noise is also incorporated. 

Following the analysis presented in Section~\ref{sec:3}, for each energy band, we combine these seven light curves to generate the PDS. Similar to the observational findings presented in Figure~\ref{fig:4}, the combined power spectra of the simulated light curves in the high-energy bands are dominant with Poisson noise, resulting in QPO peaks that are too weak to be robustly detected (see Figure~\ref{fig:B5}). To rigorously evaluate the phase-folding profiles under the impact of Poisson noise, we independently simulate $10^4$ light curves for each of the two energy bands. Then, the phase-folding is performed on the simulated total and QPO-subtracted light curves for both energy bands. It should be noted that, due to the low SNR in the high-energy band, conducting the HHT analysis there is not feasible. Consequently, the phase-folding of the high-energy light curves is carried out using the instantaneous phase obtained from the HHT analysis in the low-energy band. 

\begin{figure*}
\centering
    \begin{minipage}[c]{0.45\textwidth}
\centering
    \includegraphics[width=\linewidth]{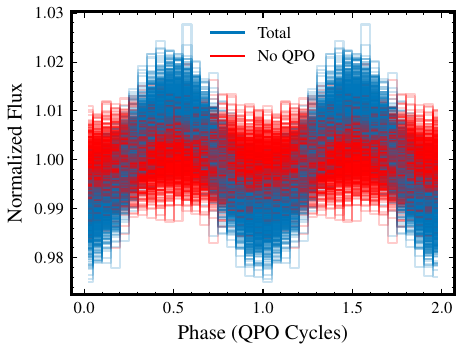}
\end{minipage}
\begin{minipage}[c]{0.45\textwidth}
\centering
    \includegraphics[width=\linewidth]{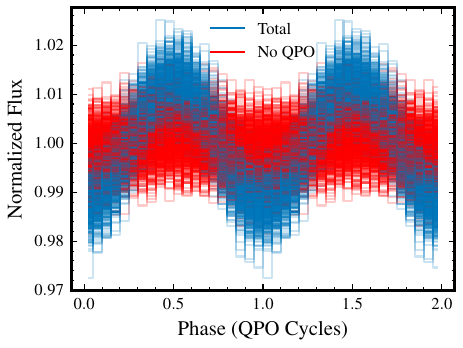}
\end{minipage}\\
\centering
    \begin{minipage}[c]{0.45\textwidth}
\centering
    \includegraphics[width=\linewidth]{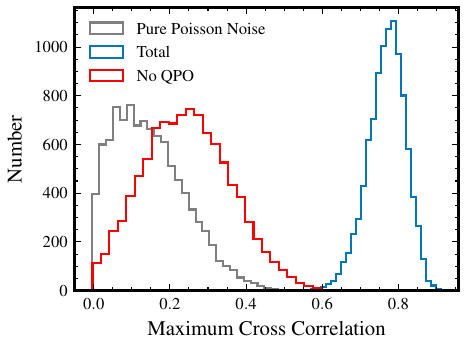}
\end{minipage}
\centering
    \begin{minipage}[c]{0.45\textwidth}
\centering
    \includegraphics[width=\linewidth]{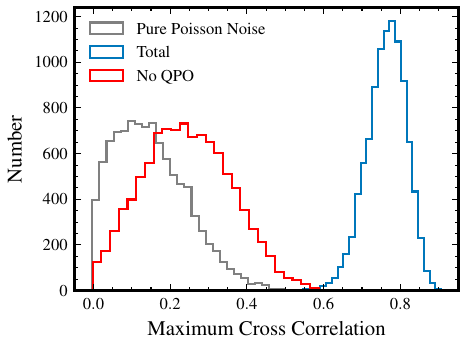}
\end{minipage}
    \caption{The top two panels respectively present phase-folding profiles of total (in blue) and QPO-subtracted (in red) light curves sampled from 10$^4$ simulated light curves in the energy ranges of 150--170 and 170--200 keV. The bottom two panels present the distributions of the maximum cross-correlations between phase-folding profiles of the simulated total (in blue), QPO-subtracted (in red), pure Poisson noise (in grey) light curves in the two high energy bands and the profile of the total light curve in the low energy band presented in Figure~\ref{fig:B4}.} \label{fig:B6}
\end{figure*}

The upper panels of Figure~\ref{fig:B6} present phase-folding profiles using 1000 samples from the simulated light curves in the two energy bands. It is evident that in both energy bands, the phase-folding profiles of the total light curves exhibit significant modulations, while those of the QPO-subtracted light curves are relatively weaker, likely being dominated by the random fluctuations of Poisson noise. Based on the cross-correlation technique described in Appendix~\ref{appenix}, for each sample, we calculate the maximum cross-correlations between phase-folding profiles of the total, QPO-subtracted, pure Poisson noise light curves in the high-energy bands and the reference profile. The reference profile is the phase-folded light curve in the low-energy band, plotted in blue in Figure~\ref{fig:B4}.

As shown in the bottom panels of Figure~\ref{fig:B6}, the distributions of maximum cross-correlations of the pure Poisson noise and QPO-subtracted signals exhibit significant overlap in both energy bands. This demonstrates the QPO-subtracted signals in the two energy bands are dominated by the Poisson noise. Therefore, the modulations in the phase-folding profiles of the QPO-subtracted signals in the high-energy bands could only be determined at the confidence level of $<2\sigma$, which is inconsistent with our observational results ($\sim$6--8 $\sigma$). Moreover, in a real case, the red noise light curves from two energy bands are not perfectly correlated, indicating that the modulations of the QPO-subtracted signals in the high energy bands could be even weaker than the simulated ones. However, the distributions of the total signals (with the QPO component) are significantly different from those of the pure Poisson and QPO-subtracted signals, indicating that the modulations of total light curves in the high energy bands could be detected at the $\sim$5--8 $\sigma$ confidence level, consistent with the observational results. 

\begin{figure*}
\centering
    \begin{minipage}[c]{0.45\textwidth}
\centering
    \includegraphics[width=\linewidth]{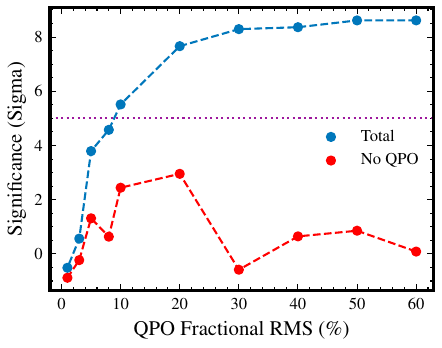}
\end{minipage}
\begin{minipage}[c]{0.45\textwidth}
\centering
    \includegraphics[width=\linewidth]{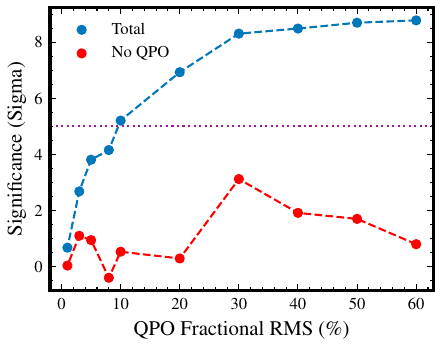}
\end{minipage}\\
\centering
    \begin{minipage}[c]{0.45\textwidth}
\centering
    \includegraphics[width=\linewidth]{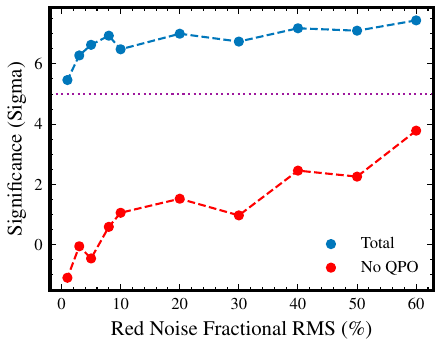}
\end{minipage}
\centering
    \begin{minipage}[c]{0.45\textwidth}
\centering
    \includegraphics[width=\linewidth]{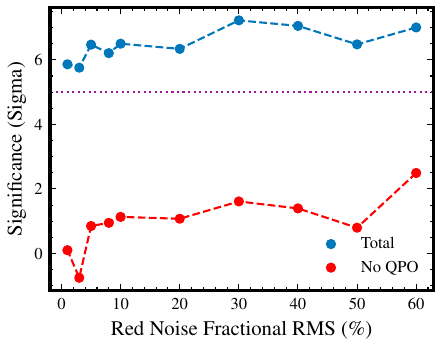}
\end{minipage}
    \caption{The top two panels respectively present relationships between QPO fractional rms and significance of modulations in the total (in blue) and QPO-subtracted (in red) profiles in the energy ranges of 150--170 and 170--200 keV, where the red noise fractional rms is fixed at 30\%. The bottom two panels present relationships between red noise fractional rms and significance of modulations in the total and QPO-subtracted profiles, where the QPO fractional rms is fixed at 15\%. The purple dotted lines plotted in four panels represented that the modulation in profiles can be determined at a $5\sigma$ confidence level.} \label{fig:B_rms}
\end{figure*}

In the simulation analysis above, we assume that the fractional rms of QPO and red noise does not vary with photon energy. However, in reality, the fractional rms of both signals may indeed change with photon energy. Therefore, we conduct tests to examine how the amplitude of the QPO and red noise affects the phase-folding method performed in the 150--170 keV and 170--200 keV energy bands. Firstly, we keep the fractional rms of the red noise constant at the value observed in the 30–120 keV energy band, while varying the QPO fractional rms. The top two panels of Figure~\ref{fig:B_rms} present the dependence of the modulation significance on the QPO rms. The significance is determined with the Monte Carlo method described in Appendix~\ref{appenix}. It is evident that the modulation significance of the total profiles increases with the QPO rms, whereas that of the QPO-subtracted profiles remain relatively constant (0--4$\sigma$). When the QPO fractional rms is lower than $\sim8\%$, the modulation of the total profiles cannot be robustly determined ($<5\sigma$). Subsequently, we fix the fractional rms of the QPO signal at the value of that in the 30--120 keV energy band, and vary the fractional rms of the red noise. As shown in the two bottom panels of Figure~\ref{fig:B_rms}, both the total and QPO-subtracted profiles exhibit modulations that can be determined with higher confidence levels when the fractional rms of the red noise is higher. However, for the cases that the fractional rms of the red noise is lower than 60\%, the modulations of the QPO-subtracted profiles could only be determined at the confidence level of $<5\sigma$, which is inconsistent with our observational results ($\sim$6--8 $\sigma$). Moreover, previous studies on the red noise in BHXRBs have indicated that its fractional rms generally decreases with photon energy \citep[see e.g.][]{2021ApJ...919...92B,2022ApJ...932....7Y}. This suggests that the fractional rms of the red noise is likely to be lower than 30\% in the 150--170 keV and 170--200 keV energy bands, and could not significantly influence the determination of the intrinsic QPO modulation.

In our simulation analysis, we model the red noise component using a broken power law function. We have also conducted the simulation analysis using the zero-frequency Lorentzian function to model the red noise, and obtained similar results. Our simulation analyses provide supports for the idea that the observed modulations of the phase-folding profile in the energy ranges of 150--170 and 170--200 keV could indeed originate from the QPO component, despite the presence of some mode mixing issues in the VMD.

\section{Determine the high-energy QPO waveforms}
\label{appenix}
\begin{figure*}
\centering
    \begin{minipage}[c]{0.45\textwidth}
\centering
    \includegraphics[width=\linewidth]{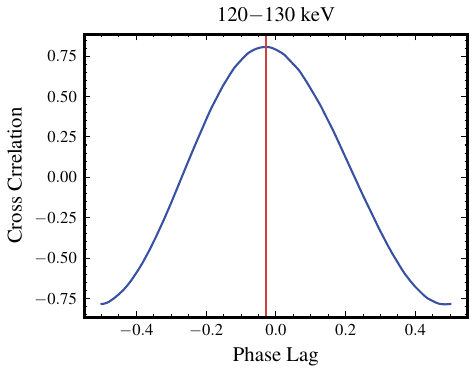}
\end{minipage}
\begin{minipage}[c]{0.45\textwidth}
\centering
    \includegraphics[width=\linewidth]{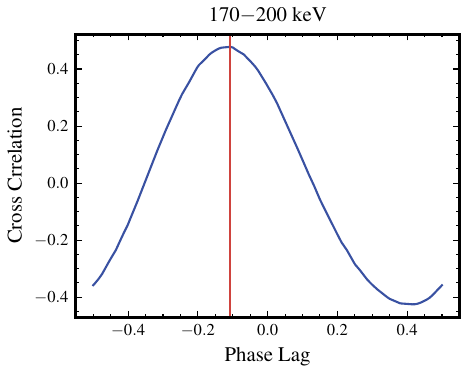}
\end{minipage}\\
\centering
    \begin{minipage}[c]{0.45\textwidth}
\centering
    \includegraphics[width=\linewidth]{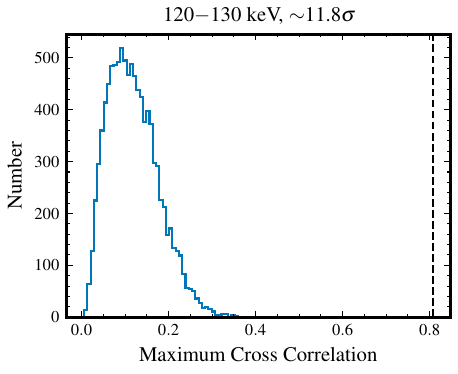}
\end{minipage}
\centering
    \begin{minipage}[c]{0.45\textwidth}
\centering
    \includegraphics[width=\linewidth]{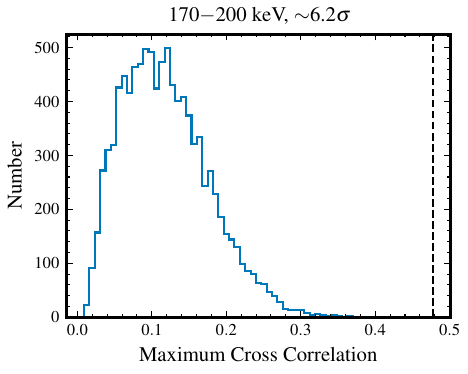}
\end{minipage}
    \caption{Examples of cross correlation of 120--130 and 170--200 keV QPO waveforms with the 30--120 keV QPO waveform. The upper panels plot the cross correlation vs. phase lag for a specified waveform with the 30--120 keV QPO waveform. The lower panels show the cross-correlation distribution from $10^4$ Monte Carlo simulations of two non-correlated waveforms. The dashed vertical lines plotted in the lower panels indicate the peak cross-correlation values as observed in the upper panels.}
    \label{fig:axp}
\end{figure*}

In this section, we present the determination of the high-energy QPO waveforms using a cross-correlation technique \citep[see][for more details]{2022ApJ...938..149H}. In a brief, this method first calculates the cross correlation of the QPO waveform in a given energy band with that in the 30–120 keV energy range, and then uses a Monte Carlo method to simulate the cross-correlation distribution between two noncorrelated Poisson sampled profiles. Appendix Figure~\ref{fig:axp} shows the cross correlation of QPO waveforms and the distribution of the cross correlation from simulated waweforms for energy ranges of 120--130 and 170--200 keV as examples. The number of bins per period is chosen as 60, but we have tested different binning of 30, 40, 50, 70, all resulting in similar significance of the QPO waveform in a given energy band. Using this technique, we detect the high-energy QPO waveform above 170 keV up to 200 keV with a significance of 6.2$\sigma$.


\bibliography{sample631}{}

\begin{thebibliography}{}
\expandafter\ifx\csname natexlab\endcsname\relax\def\natexlab#1{#1}\fi
\providecommand{\url}[1]{\href{#1}{#1}}
\providecommand{\dodoi}[1]{doi:~\href{http://doi.org/#1}{\nolinkurl{#1}}}
\providecommand{\doeprint}[1]{\href{http://ascl.net/#1}{\nolinkurl{http://ascl.net/#1}}}
\providecommand{\doarXiv}[1]{\href{https://arxiv.org/abs/#1}{\nolinkurl{https://arxiv.org/abs/#1}}}

\bibitem[{{Ackermann} {et~al.}(2015){Ackermann}, {Ajello}, {Albert}, {Atwood}, {Baldini}, {Ballet}, {Barbiellini}, {Bastieri}, {Becerra Gonzalez}, {Bellazzini}, {Bissaldi}, {Blandford}, {Bloom}, {Bonino}, {Bottacini}, {Bregeon}, {Bruel}, {Buehler}, {Buson}, {Caliandro}, {Cameron}, {Caputo}, {Caragiulo}, {Caraveo}, {Cavazzuti}, {Cecchi}, {Chekhtman}, {Chiang}, {Chiaro}, {Ciprini}, {Cohen-Tanugi}, {Conrad}, {Cutini}, {D'Ammando}, {de Angelis}, {de Palma}, {Desiante}, {Di Venere}, {Dom{\'\i}nguez}, {Drell}, {Favuzzi}, {Fegan}, {Ferrara}, {Focke}, {Fuhrmann}, {Fukazawa}, {Fusco}, {Gargano}, {Gasparrini}, {Giglietto}, {Giommi}, {Giordano}, {Giroletti}, {Godfrey}, {Green}, {Grenier}, {Grove}, {Guiriec}, {Harding}, {Hays}, {Hewitt}, {Hill}, {Horan}, {Jogler}, {J{\'o}hannesson}, {Johnson}, {Kamae}, {Kuss}, {Larsson}, {Latronico}, {Li}, {Li}, {Longo}, {Loparco}, {Lott}, {Lovellette}, {Lubrano}, {Magill}, {Maldera}, {Manfreda}, {Max-Moerbeck}, {Mayer}, {Mazziotta}, {McEnery}, {Michelson}, {Mizuno}, {Monzani},
  {Morselli}, {Moskalenko}, {Murgia}, {Nuss}, {Ohno}, {Ohsugi}, {Ojha}, {Omodei}, {Orlando}, {Ormes}, {Paneque}, {Pearson}, {Perkins}, {Perri}, {Pesce-Rollins}, {Petrosian}, {Piron}, {Pivato}, {Porter}, {Rain{\`o}}, {Rando}, {Razzano}, {Readhead}, {Reimer}, {Reimer}, {Schulz}, {Sgr{\`o}}, {Siskind}, {Spada}, {Spandre}, {Spinelli}, {Suson}, {Takahashi}, {Thayer}, {Thompson}, {Tibaldo}, {Torres}, {Tosti}, {Troja}, {Uchiyama}, {Vianello}, {Wood}, {Wood}, {Zimmer}, {Berdyugin}, {Corbet}, {Hovatta}, {Lindfors}, {Nilsson}, {Reinthal}, {Sillanp{\"a}{\"a}}, {Stamerra}, {Takalo}, \& {Valtonen}}]{2015ApJ...813L..41A}
{Ackermann}, M., {Ajello}, M., {Albert}, A., {et~al.} 2015, \apjl, 813, L41, \dodoi{10.1088/2041-8205/813/2/L41}

\bibitem[{{Axelsson} \& {Done}(2016)}]{2016MNRAS.458.1778A}
{Axelsson}, M., \& {Done}, C. 2016, \mnras, 458, 1778, \dodoi{10.1093/mnras/stw464}

\bibitem[{{Axelsson} {et~al.}(2014){Axelsson}, {Done}, \& {Hjalmarsdotter}}]{2014MNRAS.438..657A}
{Axelsson}, M., {Done}, C., \& {Hjalmarsdotter}, L. 2014, \mnras, 438, 657, \dodoi{10.1093/mnras/stt2236}

\bibitem[{{Baglio} {et~al.}(2018){Baglio}, {Russell}, {Casella}, {Noori}, {Yazeedi}, {Belloni}, {Buckley}, {Cadolle Bel}, {Ceccobello}, {Corbel}, {Coti Zelati}, {D{\'\i}az Trigo}, {Fender}, {Gallo}, {Gandhi}, {Homan}, {Koljonen}, {Lewis}, {Maccarone}, {Malzac}, {Markoff}, {Miller-Jones}, {O'Brien}, {Russell}, {Saikia}, {Shahbaz}, {Sivakoff}, {Soria}, {Testa}, {Tetarenko}, {van den Ancker}, \& {Vincentelli}}]{2018ApJ...867..114B}
{Baglio}, M.~C., {Russell}, D.~M., {Casella}, P., {et~al.} 2018, \apj, 867, 114, \dodoi{10.3847/1538-4357/aae532}

\bibitem[{{Banerjee} {et~al.}(2023){Banerjee}, {Sharma}, {Mandal}, {Das}, {Bhatta}, \& {Bose}}]{2023MNRAS.523L..52B}
{Banerjee}, A., {Sharma}, A., {Mandal}, A., {et~al.} 2023, \mnras, 523, L52, \dodoi{10.1093/mnrasl/slad057}

\bibitem[{{Bellavita} {et~al.}(2022){Bellavita}, {Garc{\'\i}a}, {M{\'e}ndez}, \& {Karpouzas}}]{2022MNRAS.515.2099B}
{Bellavita}, C., {Garc{\'\i}a}, F., {M{\'e}ndez}, M., \& {Karpouzas}, K. 2022, \mnras, 515, 2099, \dodoi{10.1093/mnras/stac1922}

\bibitem[{{Belloni} {et~al.}(2005){Belloni}, {Homan}, {Casella}, {van der Klis}, {Nespoli}, {Lewin}, {Miller}, \& {M{\'e}ndez}}]{2005A&A...440..207B}
{Belloni}, T., {Homan}, J., {Casella}, P., {et~al.} 2005, \aap, 440, 207, \dodoi{10.1051/0004-6361:20042457}

\bibitem[{{Belloni} {et~al.}(2002){Belloni}, {Psaltis}, \& {van der Klis}}]{2002ApJ...572..392B}
{Belloni}, T., {Psaltis}, D., \& {van der Klis}, M. 2002, \apj, 572, 392, \dodoi{10.1086/340290}

\bibitem[{{Belloni} \& {Stella}(2014)}]{2014SSRv..183...43B}
{Belloni}, T.~M., \& {Stella}, L. 2014, \ssr, 183, 43, \dodoi{10.1007/s11214-014-0076-0}

\bibitem[{{Bhargava} {et~al.}(2019){Bhargava}, {Belloni}, {Bhattacharya}, \& {Misra}}]{2019MNRAS.488..720B}
{Bhargava}, Y., {Belloni}, T., {Bhattacharya}, D., \& {Misra}, R. 2019, \mnras, 488, 720, \dodoi{10.1093/mnras/stz1774}

\bibitem[{{Bu} {et~al.}(2021){Bu}, {Zhang}, {Santangelo}, {Belloni}, {Zhang}, {Qu}, {Tao}, {Huang}, {Ma}, {Li}, {Zhang}, {Chen}, {Cai}, {Cao}, {Chang}, {Chen}, {Chen}, {Chen}, {Cui}, {Du}, {Gao}, {Gao}, {Ge}, {Gu}, {Guan}, {Guo}, {Han}, {Huo}, {Jia}, {Jiang}, {Jin}, {Kong}, {Li}, {Li}, {Li}, {Li}, {Li}, {Li}, {Li}, {Li}, {Li}, {Liang}, {Liao}, {Liu}, {Liu}, {Liu}, {Liu}, {Lu}, {Lu}, {Luo}, {Luo}, {Ma}, {Meng}, {Nang}, {Nie}, {Ou}, {Sai}, {Song}, {Song}, {Sun}, {Tan}, {Tuo}, {Wang}, {Wang}, {Wang}, {Wang}, {Wang}, {Wen}, {Wu}, {Wu}, {Wu}, {Xiao}, {Xiao}, {Xiong}, {Xu}, {Yang}, {Yang}, {Yi}, {Yin}, {You}, {Zhang}, {Zhang}, {Zhang}, {Zhang}, {Zhang}, {Zhang}, {Zhang}, {Zhang}, {Zhao}, {Zhao}, {Zheng}, {Zhou}, \& {Insight-HMXT Collaboration}}]{2021ApJ...919...92B}
{Bu}, Q.~C., {Zhang}, S.~N., {Santangelo}, A., {et~al.} 2021, \apj, 919, 92, \dodoi{10.3847/1538-4357/ac11f5}

\bibitem[{{Camp} {et~al.}(2007){Camp}, {Cannizzo}, \& {Numata}}]{2007PhRvD..75f1101C}
{Camp}, J.~B., {Cannizzo}, J.~K., \& {Numata}, K. 2007, \prd, 75, 061101, \dodoi{10.1103/PhysRevD.75.061101}

\bibitem[{{Cao} {et~al.}(2020){Cao}, {Jiang}, {Meng}, {Zhang}, {Luo}, {Yang}, {Zhang}, {Gu}, {Sun}, {Liu}, {Yang}, {Li}, {Tan}, {Liu}, {Du}, {Lu}, {Xu}, {Guan}, {Zhang}, {Wang}, {Li}, {Zhang}, {Wen}, {Qu}, {Song}, {Li}, {Ge}, {Zhou}, {Xiong}, {Zhang}, {Zhang}, {Cheng}, {Zhang}, {Li}, {Liang}, {Gao}, {Yang}, {Liu}, {Liu}, {Yang}, \& {Zhang}}]{2020SCPMA..6349504C}
{Cao}, X., {Jiang}, W., {Meng}, B., {et~al.} 2020, Science China Physics, Mechanics, and Astronomy, 63, 249504, \dodoi{10.1007/s11433-019-1506-1}

\bibitem[{Carvalho {et~al.}(2020)Carvalho, Moraes, Braga, \& Mendes}]{CARVALHO2020102073}
Carvalho, V.~R., Moraes, M.~F., Braga, A.~P., \& Mendes, E.~M. 2020, Biomedical Signal Processing and Control, 62, 102073, \dodoi{https://doi.org/10.1016/j.bspc.2020.102073}

\bibitem[{{Casella} {et~al.}(2005){Casella}, {Belloni}, \& {Stella}}]{2005ApJ...629..403C}
{Casella}, P., {Belloni}, T., \& {Stella}, L. 2005, \apj, 629, 403, \dodoi{10.1086/431174}

\bibitem[{{Chauhan} {et~al.}(2019){Chauhan}, {Miller-Jones}, {Anderson}, {Raja}, {Bahramian}, {Hotan}, {Indermuehle}, {Whiting}, {Allison}, {Anderson}, {Bunton}, {Koribalski}, \& {Mahony}}]{2019MNRAS.488L.129C}
{Chauhan}, J., {Miller-Jones}, J.~C.~A., {Anderson}, G.~E., {et~al.} 2019, \mnras, 488, L129, \dodoi{10.1093/mnrasl/slz113}

\bibitem[{{Chen} {et~al.}(2020){Chen}, {Cui}, {Li}, {Wang}, {Xu}, {Lu}, {Wang}, {Chen}, {Han}, {Hu}, {Zhang}, {Huo}, {Yang}, {Li}, {Lu}, {Zhang}, {Li}, {Zhang}, {Xiong}, {Zhang}, {Xue}, {Zhao}, {Zhu}, {Zhu}, {Liu}, {Yang}, \& {Zhang}}]{2020SCPMA..6349505C}
{Chen}, Y., {Cui}, W., {Li}, W., {et~al.} 2020, Science China Physics, Mechanics, and Astronomy, 63, 249505, \dodoi{10.1007/s11433-019-1469-5}

\bibitem[{{de Ruiter} {et~al.}(2019){de Ruiter}, {van den Eijnden}, {Ingram}, \& {Uttley}}]{2019MNRAS.485.3834D}
{de Ruiter}, I., {van den Eijnden}, J., {Ingram}, A., \& {Uttley}, P. 2019, \mnras, 485, 3834, \dodoi{10.1093/mnras/stz665}

\bibitem[{{Done} {et~al.}(2007){Done}, {Gierli{\'n}ski}, \& {Kubota}}]{2007A&ARv..15....1D}
{Done}, C., {Gierli{\'n}ski}, M., \& {Kubota}, A. 2007, \aapr, 15, 1, \dodoi{10.1007/s00159-007-0006-1}

\bibitem[{Dragomiretskiy \& Zosso(2014)}]{6655981}
Dragomiretskiy, K., \& Zosso, D. 2014, IEEE Transactions on Signal Processing, 62, 531, \dodoi{10.1109/TSP.2013.2288675}

\bibitem[{{Guo} {et~al.}(2020){Guo}, {Liao}, {Zhang}, {Zhang}, {Tan}, {Song}, {Lu}, {Cao}, {Chang}, {Chen}, {Du}, {Ge}, {Gu}, {Jiang}, {Jin}, {Li}, {Li}, {Li}, {Liu}, {Liu}, {Lu}, {Luo}, {Meng}, {Sun}, {Yang}, {Yang}, {You}, {Zhang}, {Zhao}, \& {Zhang}}]{2020JHEAp..27...44G}
{Guo}, C.-C., {Liao}, J.-Y., {Zhang}, S., {et~al.} 2020, Journal of High Energy Astrophysics, 27, 44, \dodoi{10.1016/j.jheap.2020.02.008}

\bibitem[{{Homan} \& {Belloni}(2005)}]{2005Ap&SS.300..107H}
{Homan}, J., \& {Belloni}, T. 2005, \apss, 300, 107, \dodoi{10.1007/s10509-005-1197-4}

\bibitem[{{Homan} {et~al.}(2020){Homan}, {Bright}, {Motta}, {Altamirano}, {Arzoumanian}, {Basak}, {Belloni}, {Cackett}, {Fender}, {Gendreau}, {Kara}, {Pasham}, {Remillard}, {Steiner}, {Stevens}, \& {Uttley}}]{2020ApJ...891L..29H}
{Homan}, J., {Bright}, J., {Motta}, S.~E., {et~al.} 2020, \apjl, 891, L29, \dodoi{10.3847/2041-8213/ab7932}

\bibitem[{{Hou} {et~al.}(2022){Hou}, {Ge}, {Ji}, {Zhang}, {You}, {Tao}, {Zhang}, {Soria}, {Feng}, {Zhou}, {Tuo}, {Song}, \& {Wang}}]{2022ApJ...938..149H}
{Hou}, X., {Ge}, M.~Y., {Ji}, L., {et~al.} 2022, \apj, 938, 149, \dodoi{10.3847/1538-4357/ac8c93}

\bibitem[{{Hsieh} \& {Chou}(2020)}]{2020ApJ...900..116H}
{Hsieh}, H.-E., \& {Chou}, Y. 2020, \apj, 900, 116, \dodoi{10.3847/1538-4357/abacbd}

\bibitem[{{Hu} {et~al.}(2011){Hu}, {Chou}, {Wu}, {Yang}, \& {Su}}]{2011ApJ...740...67H}
{Hu}, C.-P., {Chou}, Y., {Wu}, M.-C., {Yang}, T.-C., \& {Su}, Y.-H. 2011, \apj, 740, 67, \dodoi{10.1088/0004-637X/740/2/67}

\bibitem[{{Hu} {et~al.}(2014){Hu}, {Chou}, {Yang}, \& {Su}}]{2014ApJ...788...31H}
{Hu}, C.-P., {Chou}, Y., {Yang}, T.-C., \& {Su}, Y.-H. 2014, \apj, 788, 31, \dodoi{10.1088/0004-637X/788/1/31}

\bibitem[{{Huang} \& {Wu}(2008)}]{2008RvGeo..46.2006H}
{Huang}, N.~E., \& {Wu}, Z. 2008, Reviews of Geophysics, 46, RG2006, \dodoi{10.1029/2007RG000228}

\bibitem[{{Huang} {et~al.}(1998){Huang}, {Shen}, {Long}, {Wu}, {Shih}, {Zheng}, {Yen}, {Tung}, \& {Liu}}]{1998RSPSA.454..903H}
{Huang}, N.~E., {Shen}, Z., {Long}, S.~R., {et~al.} 1998, Proceedings of the Royal Society of London Series A, 454, 903, \dodoi{10.1098/rspa.1998.0193}

\bibitem[{{Huang} {et~al.}(2018){Huang}, {Qu}, {Zhang}, {Bu}, {Chen}, {Tao}, {Zhang}, {Lu}, {Li}, {Song}, {Xu}, {Cao}, {Chen}, {Liu}, {Chang}, {Yu}, {Weng}, {Hou}, {Kong}, {Xie}, {Zhang}, {ZHOU}, {Chang}, {Chen}, {Chen}, {Chen}, {Chen}, {Cui}, {Cui}, {Deng}, {Dong}, {Du}, {Fu}, {Gao}, {Gao}, {Gao}, {Ge}, {Gu}, {Guan}, {Gungor}, {Guo}, {Han}, {Hu}, {Huo}, {Ji}, {Jia}, {Jiang}, {Jiang}, {Jin}, {Jin}, {Li}, {Li}, {Li}, {Li}, {Li}, {Li}, {Li}, {Li}, {Li}, {Li}, {Li}, {Liang}, {Liao}, {Liu}, {Liu}, {Liu}, {Liu}, {Liu}, {Liu}, {Lu}, {Lu}, {Luo}, {Ma}, {Meng}, {Nang}, {Nie}, {Ou}, {Sai}, {Shang}, {Sun}, {Tan}, {Tao}, {Tuo}, {Wang}, {Wang}, {Wang}, {Wang}, {Wang}, {Wen}, {Wu}, {Wu}, {Xiao}, {Xiong}, {Xu}, {Yan}, {Yang}, {Yang}, {Yang}, {Zhang}, {Zhang}, {Zhang}, {Zhang}, {Zhang}, {Zhang}, {Zhang}, {Zhang}, {Zhang}, {Zhang}, {Zhang}, {Zhang}, {Zhang}, {Zhang}, {Zhang}, {Zhang}, {Zhang}, {Zhang}, {Zhao}, {Zhao}, {Zhao}, {Zheng}, {Zhu}, {Zhu}, {Zou}, \& {Insight-HXMT Collaboration}}]{2018ApJ...866..122H}
{Huang}, Y., {Qu}, J.~L., {Zhang}, S.~N., {et~al.} 2018, \apj, 866, 122, \dodoi{10.3847/1538-4357/aade4c}

\bibitem[{{Huppenkothen} {et~al.}(2019){Huppenkothen}, {Bachetti}, {Stevens}, {Migliari}, {Balm}, {Hammad}, {Khan}, {Mishra}, {Rashid}, {Sharma}, {Martinez Ribeiro}, \& {Valles Blanco}}]{2019ApJ...881...39H}
{Huppenkothen}, D., {Bachetti}, M., {Stevens}, A.~L., {et~al.} 2019, \apj, 881, 39, \dodoi{10.3847/1538-4357/ab258d}

\bibitem[{{Ingram} \& {Done}(2011)}]{2011MNRAS.415.2323I}
{Ingram}, A., \& {Done}, C. 2011, \mnras, 415, 2323, \dodoi{10.1111/j.1365-2966.2011.18860.x}

\bibitem[{{Ingram} {et~al.}(2009){Ingram}, {Done}, \& {Fragile}}]{2009MNRAS.397L.101I}
{Ingram}, A., {Done}, C., \& {Fragile}, P.~C. 2009, \mnras, 397, L101, \dodoi{10.1111/j.1745-3933.2009.00693.x}

\bibitem[{{Ingram} \& {van der Klis}(2015)}]{2015MNRAS.446.3516I}
{Ingram}, A., \& {van der Klis}, M. 2015, \mnras, 446, 3516, \dodoi{10.1093/mnras/stu2373}

\bibitem[{{Ingram} {et~al.}(2017){Ingram}, {van der Klis}, {Middleton}, {Altamirano}, \& {Uttley}}]{2017MNRAS.464.2979I}
{Ingram}, A., {van der Klis}, M., {Middleton}, M., {Altamirano}, D., \& {Uttley}, P. 2017, \mnras, 464, 2979, \dodoi{10.1093/mnras/stw2581}

\bibitem[{{Ingram} {et~al.}(2016){Ingram}, {van der Klis}, {Middleton}, {Done}, {Altamirano}, {Heil}, {Uttley}, \& {Axelsson}}]{2016MNRAS.461.1967I}
{Ingram}, A., {van der Klis}, M., {Middleton}, M., {et~al.} 2016, \mnras, 461, 1967, \dodoi{10.1093/mnras/stw1245}

\bibitem[{{Ingram} \& {Motta}(2019)}]{2019NewAR..8501524I}
{Ingram}, A.~R., \& {Motta}, S.~E. 2019, \nar, 85, 101524, \dodoi{10.1016/j.newar.2020.101524}

\bibitem[{{Kato}(1990)}]{1990PASJ...42...99K}
{Kato}, S. 1990, \pasj, 42, 99

\bibitem[{{Kennea} {et~al.}(2017){Kennea}, {Evans}, {Beardmore}, {Krimm}, {Romano}, {Yamaoka}, {Serino}, \& {Negoro}}]{2017ATel10700....1K}
{Kennea}, J.~A., {Evans}, P.~A., {Beardmore}, A.~P., {et~al.} 2017, The Astronomer's Telegram, 10700, 1

\bibitem[{{Kong} {et~al.}(2020){Kong}, {Zhang}, {Chen}, {Ji}, {Zhang}, {Yang}, {Tao}, {Ma}, {Qu}, {Lu}, {Bu}, {Chen}, {Song}, {Li}, {Xu}, {Cao}, {Chen}, {Liu}, {Cai}, {Chang}, {Chen}, {Chen}, {Chen}, {Cui}, {Cui}, {Deng}, {Dong}, {Du}, {Fu}, {Gao}, {Gao}, {Gao}, {Ge}, {Gu}, {Guan}, {Guo}, {Han}, {Huang}, {Huo}, {Jia}, {Jiang}, {Jiang}, {Jin}, {Li}, {Li}, {Li}, {Li}, {Li}, {Li}, {Li}, {Li}, {Li}, {Li}, {Liang}, {Liao}, {Liu}, {Liu}, {Liu}, {Liu}, {Liu}, {Liu}, {Lu}, {Lu}, {Luo}, {Luo}, {Meng}, {Nang}, {Nie}, {Ou}, {Ren}, {Sai}, {Song}, {Sun}, {Tan}, {Tuo}, {Wang}, {Wang}, {Wang}, {Wang}, {Wang}, {Wang}, {Wen}, {Wu}, {Wu}, {Wu}, {Xiao}, {Xiao}, {Xiong}, {Xu}, {Yang}, {Yang}, {Yang}, {Yi}, {You}, {Zhang}, {Zhang}, {Zhang}, {Zhang}, {Zhang}, {Zhang}, {Zhang}, {Zhang}, {Zhang}, {Zhang}, {Zhang}, {Zhang}, {Zhang}, {Zhang}, {Zhang}, {Zhang}, {Zhang}, {Zhao}, {Zhao}, {Zheng}, {Zheng}, {Zhou}, {Zhou}, {Zhu}, {Zhu}, \& {Insight-HXMT Collaboration}}]{2020JHEAp..25...29K}
{Kong}, L.~D., {Zhang}, S., {Chen}, Y.~P., {et~al.} 2020, Journal of High Energy Astrophysics, 25, 29, \dodoi{10.1016/j.jheap.2020.01.003}

\bibitem[{{Lachowicz} \& {Done}(2010)}]{2010A&A...515A..65L}
{Lachowicz}, P., \& {Done}, C. 2010, \aap, 515, A65, \dodoi{10.1051/0004-6361/200913144}

\bibitem[{{Liao} {et~al.}(2020{\natexlab{a}}){Liao}, {Zhang}, {Lu}, {Zhang}, {Li}, {Chang}, {Chen}, {Ge}, {Guo}, {Huang}, {Jin}, {Li}, {Li}, {Li}, {Liu}, {Lu}, {Nie}, {Song}, {Wang}, {You}, {Zhang}, {Zhao}, \& {Zhang}}]{2020JHEAp..27...14L}
{Liao}, J.-Y., {Zhang}, S., {Lu}, X.-F., {et~al.} 2020{\natexlab{a}}, Journal of High Energy Astrophysics, 27, 14, \dodoi{10.1016/j.jheap.2020.04.002}

\bibitem[{{Liao} {et~al.}(2020{\natexlab{b}}){Liao}, {Zhang}, {Chen}, {Zhang}, {Jin}, {Chang}, {Chen}, {Ge}, {Guo}, {Li}, {Li}, {Lu}, {Lu}, {Nie}, {Song}, {Yang}, {You}, {Zhao}, \& {Zhang}}]{2020JHEAp..27...24L}
{Liao}, J.-Y., {Zhang}, S., {Chen}, Y., {et~al.} 2020{\natexlab{b}}, Journal of High Energy Astrophysics, 27, 24, \dodoi{10.1016/j.jheap.2020.02.010}

\bibitem[{{Liska} {et~al.}(2018){Liska}, {Hesp}, {Tchekhovskoy}, {Ingram}, {van der Klis}, \& {Markoff}}]{2018MNRAS.474L..81L}
{Liska}, M., {Hesp}, C., {Tchekhovskoy}, A., {et~al.} 2018, \mnras, 474, L81, \dodoi{10.1093/mnrasl/slx174}

\bibitem[{{Liu} {et~al.}(2020){Liu}, {Zhang}, {Li}, {Lu}, {Chang}, {Li}, {Zhang}, {Jin}, {Yu}, {Zhang}, {Fu}, {Chen}, {Ji}, {Xu}, {Deng}, {Shang}, {Liu}, {Lu}, {Zhang}, {Dong}, {Li}, {Wu}, {Li}, {Wang}, {Wu}, {Zhang}, {Zhang}, {Xiong}, {Liu}, {Zhang}, {Liu}, {Yang}, \& {Zhang}}]{2020SCPMA..6349503L}
{Liu}, C., {Zhang}, Y., {Li}, X., {et~al.} 2020, Science China Physics, Mechanics, and Astronomy, 63, 249503, \dodoi{10.1007/s11433-019-1486-x}

\bibitem[{{Liu} {et~al.}(2021){Liu}, {Huang}, {Xiao}, {Bu}, {Qu}, {Zhang}, {Zhang}, {Jia}, {Lu}, {Ma}, {Tao}, {Zhang}, {Chen}, {Song}, {Li}, {Xu}, {Cao}, {Chen}, {Liu}, {Cai}, {Chang}, {Chen}, {Chen}, {Chen}, {Chen}, {Cui}, {Cui}, {Deng}, {Dong}, {Du}, {Fu}, {Gao}, {Gao}, {Gao}, {Ge}, {Gu}, {Guan}, {Guo}, {Han}, {Huo}, {Jiang}, {Jiang}, {Jin}, {Jin}, {Kong}, {Li}, {Li}, {Li}, {Li}, {Li}, {Li}, {Li}, {Li}, {Li}, {Li}, {Liang}, {Liao}, {Liu}, {Liu}, {Liu}, {Liu}, {Liu}, {Lu}, {Lu}, {Luo}, {Luo}, {Meng}, {Nang}, {Nie}, {Ou}, {Sai}, {Shang}, {Song}, {Sun}, {Tan}, {Tuo}, {Wang}, {Wang}, {Wang}, {Wang}, {Wang}, {Wang}, {Wen}, {Wu}, {Wu}, {Wu}, {Xiao}, {Xiong}, {Xu}, {Yang}, {Yang}, {Yang}, {Yang}, {Yi}, {Yin}, {You}, {Zhang}, {Zhang}, {Zhang}, {Zhang}, {Zhang}, {Zhang}, {Zhang}, {Zhang}, {Zhang}, {Zhang}, {Zhang}, {Zhang}, {Zhang}, {Zhang}, {Zhang}, {Zhang}, {Zhao}, {Zhao}, {Zheng}, {Zheng}, {Zhou}, {Zhou}, {Zhu}, {Zhuang}, \& {Zhu}}]{2021RAA....21...70L}
{Liu}, H.-X., {Huang}, Y., {Xiao}, G.-C., {et~al.} 2021, Research in Astronomy and Astrophysics, 21, 070, \dodoi{10.1088/1674-4527/21/3/70}

\bibitem[{{Ma} {et~al.}(2021){Ma}, {Tao}, {Zhang}, {Zhang}, {Bu}, {Ge}, {Chen}, {Qu}, {Zhang}, {Lu}, {Song}, {Yang}, {Yuan}, {Cai}, {Cao}, {Chang}, {Chen}, {Chen}, {Chen}, {Chen}, {Chen}, {Cui}, {Cui}, {Deng}, {Dong}, {Du}, {Fu}, {Gao}, {Gao}, {Gao}, {Gu}, {Guan}, {Guo}, {Han}, {Huang}, {Huo}, {Ji}, {Jia}, {Jiang}, {Jiang}, {Jin}, {Jin}, {Kong}, {Li}, {Li}, {Li}, {Li}, {Li}, {Li}, {Li}, {Li}, {Li}, {Li}, {Li}, {Liang}, {Liao}, {Liu}, {Liu}, {Liu}, {Liu}, {Liu}, {Liu}, {Lu}, {Lu}, {Luo}, {Luo}, {Meng}, {Nang}, {Nie}, {Ou}, {Sai}, {Shang}, {Song}, {Sun}, {Tan}, {Tuo}, {Wang}, {Wang}, {Wang}, {Wang}, {Wang}, {Wang}, {Wen}, {Wu}, {Wu}, {Wu}, {Xiao}, {Xiao}, {Xie}, {Xiong}, {Xu}, {Xu}, {Yang}, {Yang}, {Yang}, {Yi}, {Yin}, {You}, {Zhang}, {Zhang}, {Zhang}, {Zhang}, {Zhang}, {Zhang}, {Zhang}, {Zhang}, {Zhang}, {Zhang}, {Zhang}, {Zhang}, {Zhang}, {Zhang}, {Zhang}, {Zhang}, {Zhao}, {Zhao}, {Zheng}, {Zhou}, {Zhou}, {Zhu}, {Zhu}, \& {Zhuang}}]{2021NatAs...5...94M}
{Ma}, X., {Tao}, L., {Zhang}, S.-N., {et~al.} 2021, Nature Astronomy, 5, 94, \dodoi{10.1038/s41550-020-1192-2}

\bibitem[{{Ma} {et~al.}(2023){Ma}, {Zhang}, {Tao}, {Bu}, {Qu}, {Zhang}, {Zhou}, {Huang}, {Jia}, {Song}, {Zhang}, {Ge}, {Liu}, {Yang}, {Yu}, \& {Yorgancioglu}}]{2023ApJ...948..116M}
{Ma}, X., {Zhang}, L., {Tao}, L., {et~al.} 2023, \apj, 948, 116, \dodoi{10.3847/1538-4357/acc4c3}

\bibitem[{{McKinney}(2006)}]{2006MNRAS.368.1561M}
{McKinney}, J.~C. 2006, \mnras, 368, 1561, \dodoi{10.1111/j.1365-2966.2006.10256.x}

\bibitem[{{M{\'e}ndez} {et~al.}(2022){M{\'e}ndez}, {Karpouzas}, {Garc{\'\i}a}, {Zhang}, {Zhang}, {Belloni}, \& {Altamirano}}]{2022NatAs...6..577M}
{M{\'e}ndez}, M., {Karpouzas}, K., {Garc{\'\i}a}, F., {et~al.} 2022, Nature Astronomy, 6, 577, \dodoi{10.1038/s41550-022-01617-y}

\bibitem[{{Mereminskiy} {et~al.}(2018){Mereminskiy}, {Grebenev}, {Prosvetov}, \& {Semena}}]{2018AstL...44..378M}
{Mereminskiy}, I.~A., {Grebenev}, S.~A., {Prosvetov}, A.~V., \& {Semena}, A.~N. 2018, Astronomy Letters, 44, 378, \dodoi{10.1134/S106377371806004X}

\bibitem[{{Mineshige} {et~al.}(1994){Mineshige}, {Takeuchi}, \& {Nishimori}}]{1994ApJ...435L.125M}
{Mineshige}, S., {Takeuchi}, M., \& {Nishimori}, H. 1994, \apjl, 435, L125, \dodoi{10.1086/187610}

\bibitem[{{Molteni} {et~al.}(1996){Molteni}, {Sponholz}, \& {Chakrabarti}}]{1996ApJ...457..805M}
{Molteni}, D., {Sponholz}, H., \& {Chakrabarti}, S.~K. 1996, \apj, 457, 805, \dodoi{10.1086/176775}

\bibitem[{{Morgan} {et~al.}(1997){Morgan}, {Remillard}, \& {Greiner}}]{1997ApJ...482..993M}
{Morgan}, E.~H., {Remillard}, R.~A., \& {Greiner}, J. 1997, \apj, 482, 993, \dodoi{10.1086/304191}

\bibitem[{{Motta} {et~al.}(2015){Motta}, {Casella}, {Henze}, {Mu{\~n}oz-Darias}, {Sanna}, {Fender}, \& {Belloni}}]{2015MNRAS.447.2059M}
{Motta}, S.~E., {Casella}, P., {Henze}, M., {et~al.} 2015, \mnras, 447, 2059, \dodoi{10.1093/mnras/stu2579}

\bibitem[{{Nakahira} {et~al.}(2018){Nakahira}, {Shidatsu}, {Makishima}, {Ueda}, {Yamaoka}, {Mihara}, {Negoro}, {Kawase}, {Kawai}, \& {Morita}}]{2018PASJ...70...95N}
{Nakahira}, S., {Shidatsu}, M., {Makishima}, K., {et~al.} 2018, \pasj, 70, 95, \dodoi{10.1093/pasj/psy093}

\bibitem[{{Nathan} {et~al.}(2022){Nathan}, {Ingram}, {Homan}, {Huppenkothen}, {Uttley}, {van der Klis}, {Motta}, {Altamirano}, \& {Middleton}}]{2022MNRAS.511..255N}
{Nathan}, E., {Ingram}, A., {Homan}, J., {et~al.} 2022, \mnras, 511, 255, \dodoi{10.1093/mnras/stab3803}

\bibitem[{Nazari \& Sakhaei(2018)}]{7997854}
Nazari, M., \& Sakhaei, S.~M. 2018, IEEE Journal of Biomedical and Health Informatics, 22, 1059, \dodoi{10.1109/JBHI.2017.2734074}

\bibitem[{{Negoro} {et~al.}(2017){Negoro}, {Ishikawa}, {Ueno}, {Tomida}, {Sugawara}, {Isobe}, {Shimomukai}, {Mihara}, {Sugizaki}, {Serino}, {Iwakiri}, {Shidatsu}, {Matsuoka}, {Kawai}, {Sugita}, {Yoshii}, {Tachibana}, {Harita}, {Muraki}, {Morita}, {Yoshida}, {Sakamoto}, {Kawakubo}, {Kitaoka}, {Hashimoto}, {Tsunemi}, {Yoneyama}, {Nakajima}, {Kawase}, {Sakamaki}, {Ueda}, {Hori}, {Tanimoto}, {Oda}, {Tsuboi}, {Nakamura}, {Sasaki}, {Kawai}, {Yamauchi}, {Hanyu}, {Hidaka}, {Kawamuro}, \& {Yamaoka}}]{2017ATel10699....1N}
{Negoro}, H., {Ishikawa}, M., {Ueno}, S., {et~al.} 2017, The Astronomer's Telegram, 10699, 1

\bibitem[{{Nolan} {et~al.}(1981){Nolan}, {Gruber}, {Matteson}, {Peterson}, {Rothschild}, {Doty}, {Levine}, {Lewin}, \& {Primini}}]{1981ApJ...246..494N}
{Nolan}, P.~L., {Gruber}, D.~E., {Matteson}, J.~L., {et~al.} 1981, \apj, 246, 494, \dodoi{10.1086/158949}

\bibitem[{{Parikh} {et~al.}(2019){Parikh}, {Russell}, {Wijnands}, {Miller-Jones}, {Sivakoff}, \& {Tetarenko}}]{2019ApJ...878L..28P}
{Parikh}, A.~S., {Russell}, T.~D., {Wijnands}, R., {et~al.} 2019, \apjl, 878, L28, \dodoi{10.3847/2041-8213/ab2636}

\bibitem[{{Remillard} \& {McClintock}(2006)}]{2006ARA&A..44...49R}
{Remillard}, R.~A., \& {McClintock}, J.~E. 2006, \araa, 44, 49, \dodoi{10.1146/annurev.astro.44.051905.092532}

\bibitem[{{Rodi} {et~al.}(2022){Rodi}, {Jourdain}, \& {Roques}}]{2022ApJ...935...25R}
{Rodi}, J., {Jourdain}, E., \& {Roques}, J.~P. 2022, \apj, 935, 25, \dodoi{10.3847/1538-4357/ac7fff}

\bibitem[{{Russell} {et~al.}(2019){Russell}, {Tetarenko}, {Miller-Jones}, {Sivakoff}, {Parikh}, {Rapisarda}, {Wijnands}, {Corbel}, {Tremou}, {Altamirano}, {Baglio}, {Ceccobello}, {Degenaar}, {van den Eijnden}, {Fender}, {Heywood}, {Krimm}, {Lucchini}, {Markoff}, {Russell}, {Soria}, \& {Woudt}}]{2019ApJ...883..198R}
{Russell}, T.~D., {Tetarenko}, A.~J., {Miller-Jones}, J.~C.~A., {et~al.} 2019, \apj, 883, 198, \dodoi{10.3847/1538-4357/ab3d36}

\bibitem[{{Sandrinelli} {et~al.}(2014){Sandrinelli}, {Covino}, \& {Treves}}]{2014ApJ...793L...1S}
{Sandrinelli}, A., {Covino}, S., \& {Treves}, A. 2014, \apjl, 793, L1, \dodoi{10.1088/2041-8205/793/1/L1}

\bibitem[{{Scargle} {et~al.}(1993){Scargle}, {Steiman-Cameron}, {Young}, {Donoho}, {Crutchfield}, \& {Imamura}}]{1993ApJ...411L..91S}
{Scargle}, J.~D., {Steiman-Cameron}, T., {Young}, K., {et~al.} 1993, \apjl, 411, L91, \dodoi{10.1086/186920}

\bibitem[{{Shui} {et~al.}(2021){Shui}, {Yin}, {Zhang}, {Qu}, {Chen}, {Kong}, {Wang}, {Zhang}, {Song}, {Ning}, {Wang}, {Chang}, \& {Zhang}}]{2021MNRAS.508..287S}
{Shui}, Q.~C., {Yin}, H.~X., {Zhang}, S., {et~al.} 2021, \mnras, 508, 287, \dodoi{10.1093/mnras/stab2521}

\bibitem[{{Shui} {et~al.}(2023{\natexlab{a}}){Shui}, {Zhang}, {Zhang}, {Chen}, {Kong}, {Wang}, {Peng}, {Ji}, {Santangelo}, {Yin}, {Qu}, {Tao}, {Ge}, {Huang}, {Zhang}, {Liu}, {Zhang}, {Yu}, {Chang}, {Li}, {Ye}, {Li}, {Yu}, \& {Yan}}]{2023ApJ...957...84S}
{Shui}, Q.~C., {Zhang}, S., {Zhang}, S.~N., {et~al.} 2023{\natexlab{a}}, \apj, 957, 84, \dodoi{10.3847/1538-4357/acfc42}

\bibitem[{{Shui} {et~al.}(2023{\natexlab{b}}){Shui}, {Zhang}, {Chen}, {Zhang}, {Kong}, {Wang}, {Ji}, {Yin}, {Qu}, {Tao}, {Ge}, {Peng}, {Chang}, {Li}, \& {Zhang}}]{2023ApJ...943..165S}
{Shui}, Q.~C., {Zhang}, S., {Chen}, Y.~P., {et~al.} 2023{\natexlab{b}}, \apj, 943, 165, \dodoi{10.3847/1538-4357/aca7b8}

\bibitem[{{Sobolewska} \& {{\.Z}ycki}(2006)}]{2006MNRAS.370..405S}
{Sobolewska}, M.~A., \& {{\.Z}ycki}, P.~T. 2006, \mnras, 370, 405, \dodoi{10.1111/j.1365-2966.2006.10489.x}

\bibitem[{{Stella} {et~al.}(1999){Stella}, {Vietri}, \& {Morsink}}]{1999ApJ...524L..63S}
{Stella}, L., {Vietri}, M., \& {Morsink}, S.~M. 1999, \apjl, 524, L63, \dodoi{10.1086/312291}

\bibitem[{{Stevens} \& {Uttley}(2016)}]{2016MNRAS.460.2796S}
{Stevens}, A.~L., \& {Uttley}, P. 2016, \mnras, 460, 2796, \dodoi{10.1093/mnras/stw1093}

\bibitem[{{Stevens} {et~al.}(2018){Stevens}, {Uttley}, {Altamirano}, {Arzoumanian}, {Bult}, {Cackett}, {Fabian}, {Gendreau}, {Ha}, {Homan}, {Ingram}, {Kara}, {Kellogg}, {Ludlam}, {Miller}, {Neilsen}, {Pasham}, {Remillard}, {Steiner}, \& {van den Eijnden}}]{2018ApJ...865L..15S}
{Stevens}, A.~L., {Uttley}, P., {Altamirano}, D., {et~al.} 2018, \apjl, 865, L15, \dodoi{10.3847/2041-8213/aae1a4}

\bibitem[{{Stiele} \& {Kong}(2018)}]{2018ApJ...868...71S}
{Stiele}, H., \& {Kong}, A.~K.~H. 2018, \apj, 868, 71, \dodoi{10.3847/1538-4357/aae7d3}

\bibitem[{{Su} {et~al.}(2015){Su}, {Chou}, {Hu}, \& {Yang}}]{2015ApJ...815...74S}
{Su}, Y.-H., {Chou}, Y., {Hu}, C.-P., \& {Yang}, T.-C. 2015, \apj, 815, 74, \dodoi{10.1088/0004-637X/815/1/74}

\bibitem[{{Tagger} \& {Pellat}(1999)}]{1999A&A...349.1003T}
{Tagger}, M., \& {Pellat}, R. 1999, \aap, 349, 1003

\bibitem[{{Tao} {et~al.}(2018){Tao}, {Chen}, {G{\"u}ng{\"o}r}, {Huang}, {Lu}, {Qu}, {Song}, {Zhang}, {Zhang}, \& {Zhang}}]{2018MNRAS.480.4443T}
{Tao}, L., {Chen}, Y., {G{\"u}ng{\"o}r}, C., {et~al.} 2018, \mnras, 480, 4443, \dodoi{10.1093/mnras/sty2157}

\bibitem[{{Terrell}(1972)}]{1972ApJ...174L..35T}
{Terrell}, N.~James, J. 1972, \apjl, 174, L35, \dodoi{10.1086/180944}

\bibitem[{{Tetarenko} {et~al.}(2017){Tetarenko}, {Russell}, {Miller-Jones}, {Sivakoff}, \& {Jacpot Xrb Collaboration}}]{2017ATel10745....1T}
{Tetarenko}, A.~J., {Russell}, T.~D., {Miller-Jones}, J.~C.~A., {Sivakoff}, G.~R., \& {Jacpot Xrb Collaboration}. 2017, The Astronomer's Telegram, 10745, 1

\bibitem[{{Timmer} \& {Koenig}(1995)}]{1995A&A...300..707T}
{Timmer}, J., \& {Koenig}, M. 1995, \aap, 300, 707

\bibitem[{{van den Eijnden} {et~al.}(2017){van den Eijnden}, {Ingram}, {Uttley}, {Motta}, {Belloni}, \& {Gardenier}}]{2017MNRAS.464.2643V}
{van den Eijnden}, J., {Ingram}, A., {Uttley}, P., {et~al.} 2017, \mnras, 464, 2643, \dodoi{10.1093/mnras/stw2634}

\bibitem[{{van der Klis}(1989)}]{1989ARA&A..27..517V}
{van der Klis}, M. 1989, \araa, 27, 517, \dodoi{10.1146/annurev.aa.27.090189.002505}

\bibitem[{{Wagoner}(1999)}]{1999PhR...311..259W}
{Wagoner}, R.~V. 1999, \physrep, 311, 259, \dodoi{10.1016/S0370-1573(98)00104-5}

\bibitem[{Wang {et~al.}(2015)Wang, Markert, Xiang, \& Zheng}]{WANG2015243}
Wang, Y., Markert, R., Xiang, J., \& Zheng, W. 2015, Mechanical Systems and Signal Processing, 60-61, 243, \dodoi{https://doi.org/10.1016/j.ymssp.2015.02.020}

\bibitem[{{Wijnands} {et~al.}(1999){Wijnands}, {Homan}, \& {van der Klis}}]{1999ApJ...526L..33W}
{Wijnands}, R., {Homan}, J., \& {van der Klis}, M. 1999, \apjl, 526, L33, \dodoi{10.1086/312365}

\bibitem[{{Xu} {et~al.}(2018){Xu}, {Harrison}, {Garc{\'\i}a}, {Fabian}, {F{\"u}rst}, {Gandhi}, {Grefenstette}, {Madsen}, {Miller}, {Parker}, {Tomsick}, \& {Walton}}]{2018ApJ...852L..34X}
{Xu}, Y., {Harrison}, F.~A., {Garc{\'\i}a}, J.~A., {et~al.} 2018, \apjl, 852, L34, \dodoi{10.3847/2041-8213/aaa4b2}

\bibitem[{{Yang} {et~al.}(2022){Yang}, {Zhang}, {Bu}, {Huang}, {Liu}, {Yu}, {Wang}, {Tao}, {Qu}, {Zhang}, {Zhang}, {Ma}, {Song}, {Jia}, {Ge}, {Liu}, {Yan}, {Zhou}, {Li}, {Wu}, {Ren}, {Ma}, {Zhang}, {Xu}, {Ma}, {Du}, {Fu}, \& {Xiao}}]{2022ApJ...932....7Y}
{Yang}, Z.-X., {Zhang}, L., {Bu}, Q.-C., {et~al.} 2022, \apj, 932, 7, \dodoi{10.3847/1538-4357/ac63af}

\bibitem[{{Yu} {et~al.}(2023{\natexlab{a}}){Yu}, {Bu}, {Yang}, {Liu}, {Zhang}, {Huang}, {Zhou}, {Qu}, {Zhang}, {Zhang}, {Song}, {Jia}, {Ma}, {Tao}, {Ge}, {Liu}, \& {Yan}}]{2023ApJ...951..130Y}
{Yu}, W., {Bu}, Q.-C., {Yang}, Z.-X., {et~al.} 2023{\natexlab{a}}, \apj, 951, 130, \dodoi{10.3847/1538-4357/acd756}

\bibitem[{{Yu} {et~al.}(2023{\natexlab{b}}){Yu}, {Bu}, {Liu}, {Huang}, {Zhang}, {Yang}, {Qu}, {Zhang}, {Song}, {Zhang}, {Jia}, {Ma}, {Tao}, {Ge}, {Liu}, {Yan}, {Cao}, {Chang}, {Chen}, {Chen}, {Chen}, {Ding}, {Guan}, {Jin}, {Kong}, {Li}, {Li}, {Li}, {Li}, {Liao}, {Liu}, {Liu}, {Lu}, {Ma}, {Nie}, {Ren}, {Sai}, {Tan}, {Tuo}, {Wang}, {Wang}, {Wu}, {Xiao}, {Yin}, {You}, {Zhang}, {Zhang}, {Zhang}, {Zhao}, {Zheng}, \& {Zhou}}]{2023ApJ...953..191Y}
{Yu}, W., {Bu}, Q.-C., {Liu}, H.-X., {et~al.} 2023{\natexlab{b}}, \apj, 953, 191, \dodoi{10.3847/1538-4357/acd9a2}

\bibitem[{{Zhang} {et~al.}(2017){Zhang}, {Wang}, {M{\'e}ndez}, {Chen}, {Qu}, {Altamirano}, \& {Belloni}}]{2017ApJ...845..143Z}
{Zhang}, L., {Wang}, Y., {M{\'e}ndez}, M., {et~al.} 2017, \apj, 845, 143, \dodoi{10.3847/1538-4357/aa8138}

\bibitem[{{Zhang} {et~al.}(2020{\natexlab{a}}){Zhang}, {M{\'e}ndez}, {Altamirano}, {Qu}, {Chen}, {Karpouzas}, {Belloni}, {Bu}, {Huang}, {Ma}, {Tao}, \& {Wang}}]{2020MNRAS.494.1375Z}
{Zhang}, L., {M{\'e}ndez}, M., {Altamirano}, D., {et~al.} 2020{\natexlab{a}}, \mnras, 494, 1375, \dodoi{10.1093/mnras/staa797}

\bibitem[{{Zhang} {et~al.}(2023){Zhang}, {Soria}, {Zhang}, {Ji}, {Kong}, {Chen}, {Zhang}, {Chang}, {Ge}, {Li}, {Liu}, {Liu}, {Ma}, {Peng}, {Qu}, {Shui}, {Tao}, {Tian}, {Wang}, {Yan}, \& {Zeng}}]{2023A&A...677A.178Z}
{Zhang}, P., {Soria}, R., {Zhang}, S., {et~al.} 2023, \aap, 677, A178, \dodoi{10.1051/0004-6361/202346309}

\bibitem[{{Zhang} {et~al.}(2014){Zhang}, {Lu}, {Zhang}, \& {Li}}]{2014SPIE.9144E..21Z}
{Zhang}, S., {Lu}, F.~J., {Zhang}, S.~N., \& {Li}, T.~P. 2014, in Society of Photo-Optical Instrumentation Engineers (SPIE) Conference Series, Vol. 9144, Space Telescopes and Instrumentation 2014: Ultraviolet to Gamma Ray, ed. T.~{Takahashi}, J.-W.~A. {den Herder}, \& M.~{Bautz}, 914421, \dodoi{10.1117/12.2054144}

\bibitem[{{Zhang} {et~al.}(2020{\natexlab{b}}){Zhang}, {Li}, {Lu}, {Song}, {Xu}, {Liu}, {Chen}, {Cao}, {Bu}, {Chang}, {Chen}, {Chen}, {Chen}, {Chen}, {Chen}, {Cui}, {Cui}, {Deng}, {Dong}, {Du}, {Fu}, {Gao}, {Gao}, {Gao}, {Ge}, {Gu}, {Guan}, {Gungor}, {Guo}, {Han}, {Hu}, {Huang}, {Huo}, {Jia}, {Jiang}, {Jiang}, {Jin}, {Jin}, {Li}, {Li}, {Li}, {Li}, {Li}, {Li}, {Li}, {Li}, {Li}, {Li}, {Li}, {Liang}, {Liao}, {Liu}, {Liu}, {Liu}, {Liu}, {Liu}, {Liu}, {Lu}, {Lu}, {Luo}, {Ma}, {Meng}, {Nang}, {Nie}, {Ou}, {Qu}, {Sai}, {Shang}, {Shen}, {Sun}, {Tan}, {Tao}, {Tuo}, {Wang}, {Wang}, {Wang}, {Wang}, {Wang}, {Wang}, {Wang}, {Wen}, {Wu}, {Wu}, {Wu}, {Xiao}, {Xiong}, {Yan}, {Yang}, {Yang}, {Yang}, {Yi}, {Yuan}, {Zhang}, {Zhang}, {Zhang}, {Zhang}, {Zhang}, {Zhang}, {Zhang}, {Zhang}, {Zhang}, {Zhang}, {Zhang}, {Zhang}, {Zhang}, {Zhang}, {Zhang}, {Zhang}, {Zhang}, {Zhang}, {Zhang}, {Zhang}, {Zhao}, {Zhao}, {Zheng}, {Zhou}, {Zhu}, {Zhu}, {Zhuang}, \& {Insight-HXMT Team}}]{2020SCPMA..6349502Z}
{Zhang}, S.-N., {Li}, T., {Lu}, F., {et~al.} 2020{\natexlab{b}}, Science China Physics, Mechanics, and Astronomy, 63, 249502, \dodoi{10.1007/s11433-019-1432-6}

\bibitem[{{Zhang} {et~al.}(2019){Zhang}, {Ge}, {Song}, {Zhang}, {Qu}, {Zhang}, {Doroshenko}, {Tao}, {Ji}, {G{\"u}ng{\"o}r}, {Santangelo}, {Shi}, {Chang}, {Chen}, {Chen}, {Chen}, {Chen}, {Chen}, {Cui}, {Cui}, {Deng}, {Dong}, {Du}, {Fu}, {Gao}, {Gao}, {Gao}, {Gu}, {Guan}, {Guo}, {Han}, {Hu}, {Huang}, {Huo}, {Jia}, {Jiang}, {Jiang}, {Jin}, {Jin}, {Li}, {Li}, {Li}, {Li}, {Li}, {Li}, {Li}, {Li}, {Li}, {Li}, {Li}, {Liang}, {Liao}, {Liu}, {Liu}, {Liu}, {Liu}, {Liu}, {Liu}, {Liu}, {Lu}, {Lu}, {Luo}, {Ma}, {Meng}, {Nang}, {Nie}, {Ou}, {Sai}, {Sun}, {Tan}, {Tao}, {Tuo}, {Wang}, {Wang}, {Wang}, {Wang}, {Wang}, {Wen}, {Wu}, {Wu}, {Xiao}, {Xiong}, {Xu}, {Xu}, {Yan}, {Yang}, {Yang}, {Yang}, {Zhang}, {Zhang}, {Zhang}, {Zhang}, {Zhang}, {Zhang}, {Zhang}, {Zhang}, {Zhang}, {Zhang}, {Zhang}, {Zhang}, {Zhang}, {Zhang}, {Zhang}, {Zhao}, {Zhao}, {Zhao}, {Zheng}, {Zhu}, {Zhu}, {Zou}, \& {Insight-HXMT Collaboration}}]{2019ApJ...879...61Z}
{Zhang}, Y., {Ge}, M., {Song}, L., {et~al.} 2019, \apj, 879, 61, \dodoi{10.3847/1538-4357/ab22b1}

\bibitem[{{Zhang} {et~al.}(2022){Zhang}, {M{\'e}ndez}, {Garc{\'\i}a}, {Zhang}, {Karpouzas}, {Altamirano}, {Belloni}, {Qu}, {Zhang}, {Tao}, {Zhang}, {Huang}, {Kong}, {Ma}, {Yu}, {Rawat}, \& {Bellavita}}]{2022MNRAS.512.2686Z}
{Zhang}, Y., {M{\'e}ndez}, M., {Garc{\'\i}a}, F., {et~al.} 2022, \mnras, 512, 2686, \dodoi{10.1093/mnras/stac690}

\bibitem[{{Zhou} {et~al.}(2022){Zhou}, {Zhang}, {Song}, {Qu}, {Zhang}, {Ma}, {Tuo}, {Ge}, {Wang}, {Zhang}, \& {Tao}}]{2022MNRAS.515.1914Z}
{Zhou}, D.-K., {Zhang}, S.-N., {Song}, L.-M., {et~al.} 2022, \mnras, 515, 1914, \dodoi{10.1093/mnras/stac1789}

\end{thebibliography}
\bibliographystyle{aasjournal}



\end{document}